\documentclass[aps,prd,superscriptaddress,showpacs,twocolumn,nofootinbib,floatfix]{revtex4}
\usepackage{graphicx,epsfig}
\usepackage{amsmath}
\usepackage{amssymb}
\newcommand{\be}{\begin{equation}}
\newcommand{\ee}{\end{equation}}
\newcommand{\ba}{\begin{eqnarray}}
\newcommand{\ea}{\end{eqnarray}}
\newcommand{\di}{\!{\rm d}}
\newcommand{\la}{\langle}
\newcommand{\ra}{\rangle}
\newcommand{\fslash}[1] {{\not\! #1\,}}
\begin{document}
\newcommand*{\Bochum}{Institut f\"ur Theoretische Physik II,
    Ruhr-Universit\"at Bochum, D-44789 Bochum, Germany}\affiliation{\Bochum}
\newcommand*{\Coimbra}{Departamento de F\'isica and Centro de F\'isica
    Computacional, Universidade de Coimbra, P-3000 Coimbra,
    Portugal}\affiliation{\Coimbra}
\title{
    Pion mass dependence of the nucleon mass\\
    in the chiral quark soliton model}
\author{K.~Goeke}\affiliation{\Bochum}
\author{J.~Ossmann}\affiliation{\Bochum}
\author{P.~Schweitzer}\affiliation{\Bochum}
\author{A.~Silva}\affiliation{\Coimbra}

\vspace{0.5cm}

\date{December, 2005}
\begin{abstract}
The dependence of the nucleon mass on the mass of the pion is
studied in the framework of the chiral quark-soliton model.
A remarkable agreement is observed with lattice data from recent
full dynamical simulations. The possibility and limitations to use
the results from the chiral quark soliton model as a guideline
for the chiral extrapolation of lattice data are discussed.
\end{abstract}
\maketitle

\section{Introduction}
\label{Sec:1-introduction}

The formulation of QCD on a discrete, finite Euclidean lattice
\cite{Wilson:1974sk} is at present the only strict and model independent
approach allowing to solve QCD in the low energy regime and to study, e.g.,
the hadronic spectrum from first principles \cite{introduction-to-lattice}.
Numerical lattice QCD simulations face technical problems, such as
discretization errors or finite size effects, which are attacked and
minimized with increasing success by employing improved versions of
discretized actions, or by working on larger lattices available thanks
to the steadily growing  computer power. Still, present lattices are
too small to accommodate the pion as light as it appears in nature
\cite{Aoki:1999ff,AliKhan:2001tx,Bernard:2001av,Aubin:2004fs,Aoki:2002uc,Allton:2001sk,Aoki:2004ht,AliKhan:2003cu,Zanotti:2001yb}.

The tool needed to extrapolate lattice data from the region of nowadays
typically $m_\pi\gtrsim 400\,{\rm MeV}$ down to the physical value of the pion
mass is, in principle, provided by the chiral perturbation theory ($\chi$PT).
The $\chi$PT is an effective but rigorous approach to the description of low
energy phenomena of strong interactions 
\cite{Gasser:1980sb,Gasser:1983yg,Colangelo:2001df}.
It is based on the concept of spontaneous chiral symmetry breaking with
the pion as the Goldstone boson which acquires a small mass only due to
explicit chiral symmetry breaking by the small current masses of
the light quarks.
$\chi$PT allows to address such questions like, e.g., how much do
baryon masses change if one ``switches'' on the masses of light quarks
and varies their values.

In order to extrapolate reliably lattice data by means of $\chi$PT it
is important to ensure the convergence of the chiral expansion up to
large values of $m_\pi$. A first and promising {\it matching} of $\chi$PT
and lattice results was reported, and it was established that the chiral
expansion is well under control up to $m_\pi^2<0.4\,{\rm GeV}^2$
\cite{Bernard:2003rp,Procura:2003ig,Bernard:2005fy,Frink:2004ic}, see also
\cite{AliKhan:2003cu}.
More conservative estimates, however, indicate that the chiral expansion
is reliable only up to $m_\pi^2 < 0.1\,{\rm GeV}^2$ \cite{Beane:2004ks}.
The progress in computing power promisses future lattice data at still lower 
pion masses and eventually at the physical point, which will improve the 
situation and make disappear this problem. In the meantime, however,
it would be desirable to have a description of an intermediate region 
of pion masses, that would provide a safe {\it overlap} between the regime of 
the validity of $\chi$PT and the lattice data.

In this situation it is interesting to consider studies in chiral models -- 
in particular, if they allow to go beyond the range of $m_\pi$ where $\chi$PT 
is applicable. However, the inevitable prize to pay for the extended range of 
applicability compared to $\chi$PT is model-dependence, which introduces 
hardly controlable systematic uncertainties. Keeping this point critically 
in mind, such studies may nevertheless provide helpful insights. 

In Refs.~\cite{Leinweber:1998ej,Leinweber:1999ig,Leinweber:2000sa,Young:2002cj,Leinweber:2003dg}
the concept was introduced and developed to regularize chiral loops by means
of suitable vertex form factors, referred to as ``finite range regulators''
(FRR) and intended to simulate physical effects of the pion cloud which 
has a finite range due to $m_\pi\neq 0$.
As argued in \cite{Leinweber:2003dg}, the FRR method 
corresponds in some sense to a (though model-dependent)
chiral resummation reliable up to $m_\pi^2 < 1\,{\rm GeV}^2$.
While being physically intuitive and appealing, the approach was critisized 
to be unsatisfactory from a field-theoretic point of view, since it gives 
preference of one (``finite range'') regularization scheme to another 
(e.g.\ ``dimensional'') regularization scheme \cite{Bernard:2003rp}.

In this work we address the question how the nucleon mass depends (implicitly) 
on the pion mass in another effective approach, namely in the chiral 
quark soliton model ($\chi$QSM) \cite{Diakonov:yh,Diakonov:1987ty}.
This model was derived under certain assumptions from the instanton model
of the QCD vacuum \cite{Diakonov:1983hh,Diakonov:1985eg}, which provides a
dynamical picture of the chiral symmetry breaking mechanism
\cite{Diakonov:2000pa}. In this model the nucleon appears as
a soliton of the chiral pion mean field in the limit of a large number of
colours $N_c$. The model provides a theoretically consistent description
of numerous baryonic quantities ranging from static properties
\cite{Diakonov:1988mg,Christov:1995vm} over ``usual'' \cite{Diakonov:1996sr}
till ``generalized'' parton distribution functions \cite{Petrov:1998kf},
which -- as far as these quantities are known -- agree with phenomenology
to within $(10-30)\%$ {\sl at the physical point}.

Here --- focussing on the nucleon mass $M_N$ --- we present the first 
study in the $\chi$QSM at {\sl non-physical pion masses} $m_\pi$
covering the wide range $0 \le m_\pi\le 1500\,{\rm MeV}$. 
We make several remarkable observations.
First, stable soliton solutions do exist in this range of pion masses.
Second, we demonstrate that the model correctly describes also the heavy 
quark limit. 
In the opposite limit $m_\pi\to0$, which does not commute 
with large-$N_c$ limit \cite{Gasser:1980sb,Dashen:1993jt},
the $\chi$QSM is known to exhibit a chiral behaviour and to incorporate 
leading non-analytic terms, which are at variance with the real world QCD with
a finite number of colours $N_c=3$, but in agreement with its formulation 
in the limit $N_c\to\infty$ \cite{Schuren:1991sc,Cohen:1992uy,Schweitzer:2003sb}.
This is consistent as the model is defined in this limit.
Third, we show that the $\chi$QSM provides a satisfactory description of 
the ${\sl variation}$ of the lattice data on $M_N$ with $m_\pi$ in the 
considered range of pion masses.

Partly we provide explanations for these observations. Partly, however, 
they shall remain puzzles to be resolved upon further studies in the model.


This note is organized as follows.
In Sect.~\ref{Sec:2-pion-nucleon-in-eff-theory} the $\chi$QSM is introduced
and model results for $M_N(m_\pi)$ presented, which we compare to
lattice QCD in Sect.~\ref{Sec:3-compare-to-lattice}.
In Sect.~\ref{Sec:4-using-for-extrapolation?} we discuss the limitations 
for using the model quantitatively to extrapolate lattice data, and compare 
in Sect.~\ref{Sec:5-compare-to-CPT} to $\chi$PT and the FRR approach.
Sect.~\ref{Sec:5-conclusions} contains the conclusions.
Technical details and a digression on the pion-nucleon sigma-term 
can be found in the Appendices.

\section{Pion and nucleon in the effective theory}
\label{Sec:2-pion-nucleon-in-eff-theory}

Let us consider the effective theory which was derived from the instanton
model of the QCD vacuum \cite{Diakonov:1983hh,Diakonov:1985eg} and is
given by the partition function \cite{Diakonov:1984tw,Dhar:1985gh}
\be\label{eff-theory}
    Z_{\rm eff} = \int\!\!{\cal D}\psi\,{\cal D}\bar{\psi}\,
    {\cal D}U\,\exp\Biggl(i\int\di^4x\;\bar{\psi}\,
    (i\fslash{\partial}-M\,U^{\gamma_5}-m)\psi\Biggr)\,,\ee
where $U=\exp(i\tau^a\pi^a)$ denotes the $SU(2)$ chiral pion field
with $U^{\gamma_5} = \exp(i\gamma_5\tau^a\pi^a)$, and $M$ is the dynamical
(``constituent'') quark mass due to spontaneous breakdown of chiral
symmetry, and $m=m_u=m_d$ is the current quark mass.
We neglect throughout isospin breaking effects.

The effective theory (\ref{eff-theory}) is valid at low energies
below a scale set by the inverse of the average instanton size
$\rho_{\rm av}^{-1} \approx 600\,{\rm MeV}$. The dynamical mass
is momentum dependent, i.e.\ $M=M(p)$, and goes to zero for
$p\gg \rho_{\rm av}^{-1}$. In practical calculations it is convenient to
replace $M(p)$ by a constant mass $M$, and to regularize the effective
theory within an appropriate regularization scheme with a cutoff 
$\Lambda_{\rm cut}={\cal O}(\rho_{\rm av}^{-1})$. In the present work we use
$M=350\,{\rm MeV}$ from the instanton vacuum model \cite{Diakonov:2000pa}.

In the effective theory (\ref{eff-theory}) chiral symmetry is
spontaneously broken and a non-zero quark-vacuum condensate
$\la\bar\psi\psi\ra\equiv\la{\rm vac}|(\bar\psi_u\psi_u+\bar\psi_d\psi_d)|{\rm vac}\ra$
appears which is given in leading order of the large-$N_c$ limit by the
quadratically UV-divergent Euclidean loop integral
\be\label{Eq:vacuum-condensate}
    \la\bar\psi\psi\ra = - \int\frac{d^4p_E}{(2\pi)^4}\,
    \frac{8N_c{M^\prime}}{p_E^2+{M^\prime}^2}\biggr|_{reg}
        =-8N_c{M^\prime} I_1(m)\,,
\ee
where $I_1$ is its proper-time regularized version,
see App.~\ref{App:A-proper-time}, and $M^\prime \equiv M+m$.
Note that in QCD strictly speaking $\la\bar\psi\psi\ra$ is well-defined
only in the chiral limit. The pion is not a dynamical degree of freedom
in the theory (\ref{eff-theory}). Instead the dynamics of the pion field
appears only after integrating out the quark fields which yields the
effective action
\be\label{Eq:pion-action}
    S_{\rm eff}[U] = \frac{f_\pi^2}{4} \int\di^4 x \;
    {\rm tr}\;\partial^\mu U\,\partial_\mu U^\dag + \dots
\ee
where the dots denote the four-derivative Gasser-Leutwyler terms with
correct coefficients, the Wess-Zumino term, terms $\propto m$ (and an
infinite series of higher-derivative terms) \cite{Diakonov:1987ty}.
The pion decay constant $f_\pi=93\,{\rm MeV}$ in Eq.~(\ref{Eq:pion-action})
is given in the effective theory by the logarithmically UV-divergent loop
integral (whose regularized version we denote by $I_2$,
see App.~\ref{App:A-proper-time})
\be\label{Eq:fpi}
    f_\pi^2 = \int\frac{d^4 p_E}{(2\pi)^4}\;
    \frac{4N_c{M^\prime}^2}{(p_E^2+{M^\prime}^2)^2}\biggr|_{reg}
    = 8N_c{M^\prime}^2 \; I_2(m) \;.
\ee
The mass of the pion can be determined from the position of the pole of the
pion propagator in the effective theory (\ref{eff-theory}). 
Its relation to the current quark mass is given by the equation 
(for the $I_i$ see App.~\ref{App:A-proper-time})
\be\label{Eq:mpi}
    m_\pi^2 = \frac{m}{M}\;\frac{I_1(m)}{I_2(m)}\;,
\ee
which, for small current quark masses $m$, corresponds to the
Gell-Mann--Oakes--Renner relation
\be
    m_\pi^2 f_\pi^2 = -m \,\la\bar\psi\psi\ra + {\cal O}(m^2)\;.
\ee

The $\chi$QSM is an application of the effective theory (\ref{eff-theory})
to the description of baryons \cite{Diakonov:yh,Diakonov:1987ty}.
The large-$N_c$ limit allows to solve the path integral over pion field
configurations in Eq.~(\ref{eff-theory}) in the saddle-point approximation.
In the leading order of the large-$N_c$ limit the pion field is static, and
one can determine the spectrum of the one-particle Hamiltonian of the
effective theory (\ref{eff-theory})
\be\label{Hamiltonian}
    \hat{H}|n\ra=E_n |n\ra \;,\;\;
    \hat{H}=-i\gamma^0\gamma^k\partial_k+\gamma^0MU^{\gamma_5}+\gamma^0m
    \;. \ee
The spectrum consists of an upper and a lower Dirac continuum, distorted by
the pion field as compared to continua of the free Dirac-Hamiltonian
\be\label{free-Hamiltonian}
    \hat{H}_0|n_0\ra = E_{n_0}|n_0\ra \;,\;\;
    \hat{H}_0 = -i\gamma^0\gamma^k\partial_k+\gamma^0 M+\gamma^0m\, , \ee
and of a discrete bound state level of energy $E_{\rm lev}$,
if the pion field is strong enough.
By occupying the discrete level and the states of the lower continuum each
by $N_c$ quarks in an anti-symmetric colour state, one obtains a state
with unity baryon number.
The soliton energy $E_{\rm sol}$ is a functional of the pion field
\be\label{Eq:soliton-energy}
    E_{\rm sol}[U] = N_c \biggl[E_{\rm lev}+
    \sum\limits_{E_n<0}(E_n-E_{n_0})\biggr]_{\rm reg} \;.
    \ee
$E_{\rm sol}[U]$ is logarithmically divergent, see
App.~\ref{App:A-proper-time} for the explicit expression
in the proper-time regularization.
By minimizing $E_{\rm sol}[U]$ one obtains the self-consistent solitonic
pion field $U_c$.  This procedure is performed for symmetry reasons in
the so-called hedgehog ansatz
\be\label{hedgehog}
    \pi^a({\bf x})=e^a_r\;P(r) \;,\;\;
    U({\bf x})=\cos P(r)+i \tau^a e_r^a \sin P(r)\;,\ee
with the radial (soliton profile) function $P(r)$ and $r=|{\bf x}|$,
${\bf e}_r = {\bf x}/r$.
The nucleon mass $M_N$ is given by $E_{\rm sol}[U_c]$.
The self-consistent profile satisfies $P_c(0)=-\pi$ and decays in the chiral
limit as $1/r^2$ at large $r$.  For finite $m$ it exhibits a Yukawa tail
$\propto\exp(-m_\pi r)/r$ with the pion mass $m_\pi$ connected to $m$
by the relation (\ref{Eq:mpi}).

For the following discussion we note that the soliton energy can be
rewritten as an expansion in powers of $\partial^\mu U$ as follows
\be\label{Eq:gradient-expansion}
    E_{\rm sol}[U] = \sum_{k=1}^\infty F_k[(\partial U)^k]
\ee
where $F_k[(\partial U)^k]$ symbolically denotes a functional in which
$\partial^\mu U$ appears $k$-times (appropriately contracted). Note that in
leading order of the large-$N_c$ limit the soliton field is time-independent,
i.e.\ $\partial U$ is just $\nabla U$. For some observables the lowest orders
in expansions analog to (\ref{Eq:gradient-expansion}) were computed
\cite{Diakonov:1988mg,Christov:1995vm,Diakonov:1996sr,Petrov:1998kf}.

Let us also remark that in the case $m_u=m_d=m_s$ the above given formulae
for the soliton energy in the SU(2) version of the model coincide with those
from the SU(3) version \cite{Christov:1995vm}.

In order to describe further properties of the nucleon, 
it is
necessary to integrate over the zero modes of the soliton solution in the path
integral, which assigns a definite momentum, and spin and isospin quantum
numbers to the baryon. Corrections in the $1/N_c$-expansion can be included by
considering time dependent pion field fluctuations around the solitonic
solution. The $\chi$QSM allows to evaluate baryon matrix elements of local
and non-local QCD quark bilinear operators like
$\la B'|\bar{\Psi}\Gamma\Psi|B\ra$ (with $\Gamma$
denoting some Dirac- and flavour-matrix) with no adjustable parameters.
This provides the basis for the wide range of applicability of this model
\cite{Diakonov:1988mg,Christov:1995vm,Diakonov:1996sr,Petrov:1998kf}.

\subsection{The mass of the nucleon in the \boldmath large-$N_c$ limit}
\label{Subsec:Mn-in-large-Nc}

If one managed to solve QCD in the limit $N_c\to\infty$ one would in principle
obtain for the mass of the nucleon an expression of the form
(let the $M_i$ be independent of $N_c$)
\be\label{Eq:Mn-in-large-Nc}
    M_N=N_c M_1+N_c^0 M_2+N_c^{-1} M_3+\dots \;\;.
\ee
The $\chi$QSM provides a practical realization of the large-$N_c$ picture of
the nucleon \cite{Witten:tx}, and respects the general large-$N_c$ counting
rules.
In leading order of the large-$N_c$ limit one approximates in the $\chi$QSM
the nucleon mass by the expression for the soliton energy in
Eq.~(\ref{Eq:soliton-energy}), i.e.\ one considers only the first term in
the expansion (\ref{Eq:Mn-in-large-Nc}) $M_N \approx N_c  M_1 = E_{sol}$.
We obtain numerically\footnote{
    \label{footnote:accuracy}
    In this work we quote numerical results to within an accuracy of
    $1\,{\rm MeV}$. However, one should keep in mind that we neglect
    isospin breaking effects (and electromagnetic corrections).
    Therefore we shall round off the physical masses as
    $m_\pi=140\,{\rm MeV}$ and $M_N = 940\,{\rm MeV}$, and the same
    is understood for our numerical results.}
\be\label{Mn-at-mpi=140MeV}
    M_N = 1254\,{\rm MeV}
\ee
where the cutoff in the regulator function $R$, see
App.~\ref{App:A-proper-time}, is chosen such that for $m_\pi=140\,{\rm MeV}$
the physical value of the pion decay constant $f_\pi$ is reproduced.
We observe an overestimate of the physical value $M_N = 940\,{\rm MeV}$ by
about $30\%$. This is not surprising, given the fact that we truncate the
series in Eq.~(\ref{Eq:Mn-in-large-Nc}) after the first term and thus neglect
corrections which are generically of ${\cal O}(1/N_c)$, i.e. of the order
of magnitude of the observed overestimate.

In fact, the soliton approach is known to overestimate systematically the
physical values of the baryon masses because of -- among others -- spurious
contributions due to the soliton center of mass motion \cite{Pobylitsa:1992bk}.
Taking into account the corresponding corrections, which are ${\cal O}(N_c^0)$,
reduces the soliton energy by the right amount of about $300\,{\rm MeV}$.
(Note that there are also other sources of corrections at ${\cal O}(N_c^0)$,
see Ref.~\cite{Pobylitsa:1992bk}.) We shall keep in mind this systematic
overestimate, when we will discuss lattice data below.

\subsection{The chiral limit}
\label{Subsec:chiral-limit}

In the following we will be interested in particular in the pion mass
dependence of the nucleon mass. From $\chi$PT we know that
\be\label{Eq:Mn-of-mpi-in-chiPT}
    M_N(m_\pi) = M_N(0) + A \, m_\pi^2 + B\, m_\pi^3 + \dots\;\; ,
\ee
where the dots denote terms which vanish faster than $m_\pi^3$ with
$m_\pi\to 0$. The constants $M_N(0)$ and $A$ (which is related to the
pion-nucleon sigma-term) serve to absorb infinite counter-terms in the
process of renormalization in $\chi$PT -- in the sense of renormalizability
in $\chi$PT. However, the constant $B$, which accompanies the so-called
leading non-analytic (in $m$, since $m_\pi^3\propto m^{3/2}$) contribution,
is finite. For this constant $\chi$PT, as well as any theory which correctly
incorporates chiral symmetry, yields \cite{Gasser:1980sb,Dashen:1993jt}
\be\label{Eq:LNA-contribution}
    B = -\,k\;\frac{3 g_A^2}{32\pi\,f_\pi^2}\;.
\ee
with $k=1$ for finite $N_c$. However, the limits $m_\pi\to 0$ and
$N_c\to\infty$ do not commute. If we choose first to take $N_c\to\infty$
while keeping $m_\pi$ finite, and only then we consider the chiral limit,
then $k=3$. The reason for that is the special role played by the
$\Delta$-resonance. In the large-$N_c$ limit the nucleon and the
$\Delta$-resonance become mass-degenerated\footnote{
    Though in the $\chi$QSM one cannot handle ${\cal O}(N_c^0)$ corrections
    to $M_N$, still one is able to describe consistently the
    mass difference $(M_\Delta - M_N)$ and to reproduce the large-$N_c$
    counting rule in Eq.~(\ref{Eq:Mdelta-Mn}) by considering a class of
    particular (so-called ``rotational'') $1/N_c$-corrections
    \cite{Diakonov:1985eg}. In the soliton picture $\Delta$ and nucleon
    are just different rotational states of the same classical soliton.
    In this note we do not consider rotational corrections and
    work to leading-order in the large-$N_c$ expansion.}
\be\label{Eq:Mdelta-Mn}
    M_\Delta - M_N = {\cal O}(N_c^{-1})\;.
\ee
Taking $N_c\to\infty$ while $m_\pi$ is kept finite, one has
$M_\Delta-M_N\ll m_\pi$ and has to consider the contribution of the
$\Delta$-resonance as intermediate state in chiral loops on equal footing
with the contribution of the nucleon.
The contribution of the $\Delta$-resonance appears (in quantities which do not
involve polarization) to be twice the contribution of the nucleon, hence $k=3$
\cite{Cohen:1992uy}. This is the situation in the $\chi$QSM and, in fact,
in this  model one recovers \cite{Schweitzer:2003sb} the correct (in the
large-$N_c$ limit) leading non-analytic term in
Eq.~(\ref{Eq:LNA-contribution}).

\subsection{The heavy quark limit}
\label{Subsec:heavy-quark}

The $\chi$QSM is based on chiral symmetry which consideration makes sense
only when explicit chiral symmetry breaking effects are small. Nevertheless
one can consider the model, in principle, for any value of $m$. In this
context let us first note that taking the limit of a large current quark mass
{$m\to m_Q$} in Eqs.~(\ref{Eq:vacuum-condensate},~\ref{Eq:fpi},~\ref{Eq:mpi}) 
yields, see Table~\ref{Table:dependence-on-mpi}
\be\label{Eq:heavy-meson}
    \lim_{{m\to m_Q}} m_\pi = 2 \, {m_Q}
\ee
which is the correct heavy quark limit for the mass of a meson (see
Sec.\ref{Subsec-fix-para} on details of parameter fixing).

What happens to the mass of the soliton in this limit?
Consider the eigenvalue problem in Eq.~(\ref{Hamiltonian}).
With increasing $m{\to m_Q}$ the ``potential term'' $MU^{\gamma_5}$ is less and 
less important and the spectrum of the full Hamiltonian (\ref{Hamiltonian}) becomes
more and more similar to the spectrum of the free Hamiltonian
(\ref{free-Hamiltonian}). In this limit the eigenvalues of the
full and free Hamiltonian nearly cancel in the sum over energies in
Eq.~(\ref{Eq:soliton-energy}). Only the contribution of the discrete level
is not compensated and approaches $E_{\rm lev}\approx {m_Q}$.
Thus, we recover the correct heavy quark limit
\be\label{Eq:heavy-baryon}
    \lim_{{m\to m_Q}} M_N = N_c\, {m_Q} \;. \ee
However, the above considerations are formal, since the crucial step
consists in demonstrating that a stable soliton solution indeed exists, i.e.\
that a self-consistent profile can be found for which the soliton energy
(\ref{Eq:soliton-energy}) takes a minimum. In practice, we find that stable
soliton solutions exist at least up to $m={\cal O}(700\,{\rm MeV})$,
which is sufficient for a comparison to lattice QCD results. In this range
we observe a tendency to approach the limit (\ref{Eq:heavy-baryon}) from below
as expected, see Table~\ref{Table:dependence-on-mpi}. Thus, we find that both
the pion and the nucleon mass are correctly described in the effective theory
when the current mass of the quarks becomes large.

\begin{figure}[t!]
        \includegraphics[width=9.3cm,height=5.8cm]{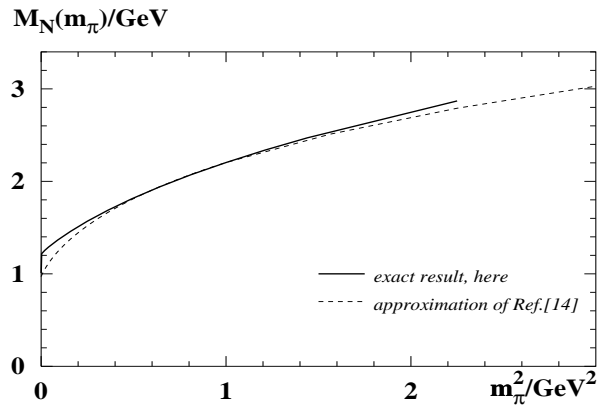}
    \caption{\footnotesize\sl
    \label{Fig-1}
    Nucleon mass $M_N(m_\pi)$ vs.\  $m_\pi^2$ in the chiral
    quark-soliton model. \underline{Solid line:} Exact result
    obtained here. \underline{Dashed line:} The approximation
    based on the instanton vacuum model from
    Ref.~\cite{Schweitzer:2003sb}.}
\end{figure}

\subsection{Fixing of parameters}
\label{Subsec-fix-para}

When studying the nucleon mass as function of $m_\pi$ we must specify
what is fixed and what varies in the chiral limit. Here we make the
following choice. We keep the dynamical constituent quark mass 
$M=350\,{\rm MeV}$ and $f_\pi=93\,{\rm MeV}$ fixed. 
In this way we obtain the results shown in
Table~\ref{Table:dependence-on-mpi}, and plotted as solid line in 
Fig.~\ref{Fig-1}. The meaning of the dashed curve in Fig.~\ref{Fig-1} is 
explained in App.~\ref{App:B-sigma} where a digression is given on the 
pion-nucleon sigma-term related to the slope of $M_N(m_\pi)$.

It should be noted that keeping $f_\pi$ fixed in the 
chiral limit is a choice often considered in literature. 
However, at this point other choices could be considered as well. 
E.g.\  one could fix the pion decay constant to its value $F=88\,{\rm MeV}$ 
in the chiral limit, or allow $f_\pi$ to be $m_\pi$-dependent, 
which strictly speaking is the case in $\chi$PT and in lattice QCD. 
In $\chi$PT 
\be\label{Eq:fpi-in-CPT}
	f_\pi(m_\pi)=F\left(1+\frac{m_\pi^2}{(4\pi F)^2}\;\bar{l}_4+
	{\cal O}\left(\frac{m_\pi^4}{F^4}\right)\right)
\ee
where $\bar{l}_4$ is a low energy constant 
\cite{Gasser:1983yg,Colangelo:2001df}.
In lattice calculations $f_\pi$ increases with larger $m_\pi$, 
exceeding its physical value by about $40\,\%$ at $m_\pi\sim 1\,{\rm GeV}$,
e.g.\ \cite{Gockeler:2005mh}. 
A more consistent way of fixing model parameters could consists in choosing 
$\Lambda_{\rm cut}$ (and/or $M$) such that in the model $f_\pi(m_\pi)$ 
satisfies (\ref{Eq:fpi-in-CPT}) with the correct value for $\bar{l}_4$, 
{\sl and} agrees in each case with lattice results at large $m_\pi$.

Remarkably, Eq.~(\ref{Eq:fpi-in-CPT}) holds in the model and parameters 
can be fixed to reproduce $\bar{l}_4$ correctly \cite{Schuren:1991sc}. 
However, it is a subtle issue, how to simulate in a chiral model
the lattice situation in a realistic way. There a common procedure is to 
keep fixed all bare lattice parameters but the bare current quark mass 
(the hopping parameter). 
In some sense, the dimensionful quantity, which is kept fixed in 
lattice calculations while $m_\pi$ ``is varied'', is the Sommer 
scale \cite{Sommer:1993ce} defined by the notion of the heavy quark 
potential --- absent in chiral models.

Notice that the $\chi$QSM describes numerous observables to within
an accuracy of typically $(10-30)\%$ \cite{Christov:1995vm}, 
but cannot be expected to be much more precise than that.
Overestimating the value of $f_\pi$ in the chiral limit 
by $5\%$ or underestimating its lattice values by $30\%$ (or $40\%$) lies 
within the accuracy of the model. 
Note that changing $f_\pi$ (at non-physical $m_\pi$) would alter 
in particular values of the cutoff $\Lambda_{\rm cut}$ and consequently 
change $M_N$ in Table~\ref{Table:dependence-on-mpi}. 
Since $f_\pi$ and $M_N$ depend on $\Lambda_{\rm cut}$ logarithmically and thus 
weakly, these changes may be expected to be small and within the accuracy of 
the model.

From a practical point of view, our choice to keep $f_\pi$ fixed to its 
physical value may be considered as {\sl one} effective prescription, 
whose consequences one may think absorbed in the unavoidable model 
dependence of the results. 

It would be interesting to consider {\sl other} effective prescriptions 
which would provide more insights into the model dependence of our study.
This is, however, beyond the scope of this work and subject to further 
investigations \cite{work-in-progress}.

%
%
\begingroup
\squeezetable
\begin{table}[ht!]
    \caption{\footnotesize\sl
    \label{Table:dependence-on-mpi}
    The dependence on the pion mass for fixed $f_\pi=93\,{\rm MeV}$
    in the $\chi$QSM. All numbers are in units of MeV.
    \underline{Rows 2,3,4:}
    Current quark mass $m$, cutoff of the effective
    theory $\Lambda_{\rm cut}$, and quark vacuum condensate
    $\la\bar\psi\psi\ra$ depend on $m_\pi$ according
    to Eqs.~(\ref{Eq:vacuum-condensate},~\ref{Eq:fpi},~\ref{Eq:mpi}).
    $\Lambda_{\rm cut}$ is of the order of magnitude of the inverse of the
    average instanton size $\rho^{-1}_{\rm av} \approx 600\,{\rm MeV}$.
    Note that in QCD -- in contrast to effective theories with a well
    defined regularization prescription -- the notion of quark vacuum
    condensate for $m\neq 0$ is ambiguous.
    \underline{Rows 5,6,7:} Contributions of the discrete level
    $N_c E_{\rm lev}$ and the continuum $N_c E_{\rm cont}$ to the total
    soliton energy $E_{\rm sol}= N_c(E_{\rm lev} + E_{\rm cont})$,
    see Eq.~(\ref{Eq:soliton-energy}),
    to be identified with the nucleon mass in the model.
    The numerical numbers confirm within the studied range of $m$ the
    heavy quark limit discussed in Sec.~\ref{Subsec:heavy-quark}.}
\vspace{0.4cm}
\begin{ruledtabular}
\begin{tabular}{ccccccc}
    \\
    $m_\pi$ &
    $m$ &
    $\Lambda_{\rm cut}$ &
    $-\la\bar\psi\psi\ra^{1/3}$ &
    $N_c E_{\rm lev}$  &
    $N_c E_{\rm cont}$ &
    $E_{\rm sol}$
    \\
    \\
    \hline
    \\
    0    &  0    & 649 & 220 &  681  & 530 & 1211 \\
    10   &  0.1  & 649 & 220 &  682  & 530 & 1212 \\
    50   &  2    & 648 & 220 &  692  & 526 & 1218 \\
    140  &  16   & 643 & 218 &  744  & 510 & 1254 \\
    300  &  69   & 635 & 211 &  872  & 494 & 1366 \\
    600  &  223  & 666 & 204 & 1202  & 484 & 1686 \\
    1200 &  556  & 799 & 205 & 2006  & 465 & 2471 \\
    \\
\end{tabular}
\end{ruledtabular}
\end{table}
\endgroup
%
%

\section{Comparison to lattice data}
\label{Sec:3-compare-to-lattice}

In this Section we compare $M_N(m_\pi)$ from $\chi$QSM with lattice results
on the nucleon mass from Refs.~\cite{AliKhan:2001tx,Bernard:2001av,Aubin:2004fs,Aoki:2002uc,Allton:2001sk,Aoki:2004ht,AliKhan:2003cu,Zanotti:2001yb}.
Wherever possible we will use only such lattice data, where all bare
lattice parameters where kept fixed apart from the current quark mass
(or the hopping paramater).

\begin{figure*}[t!]
\begin{tabular}{cc}
    \includegraphics[width=9.3cm,height=5.5cm]{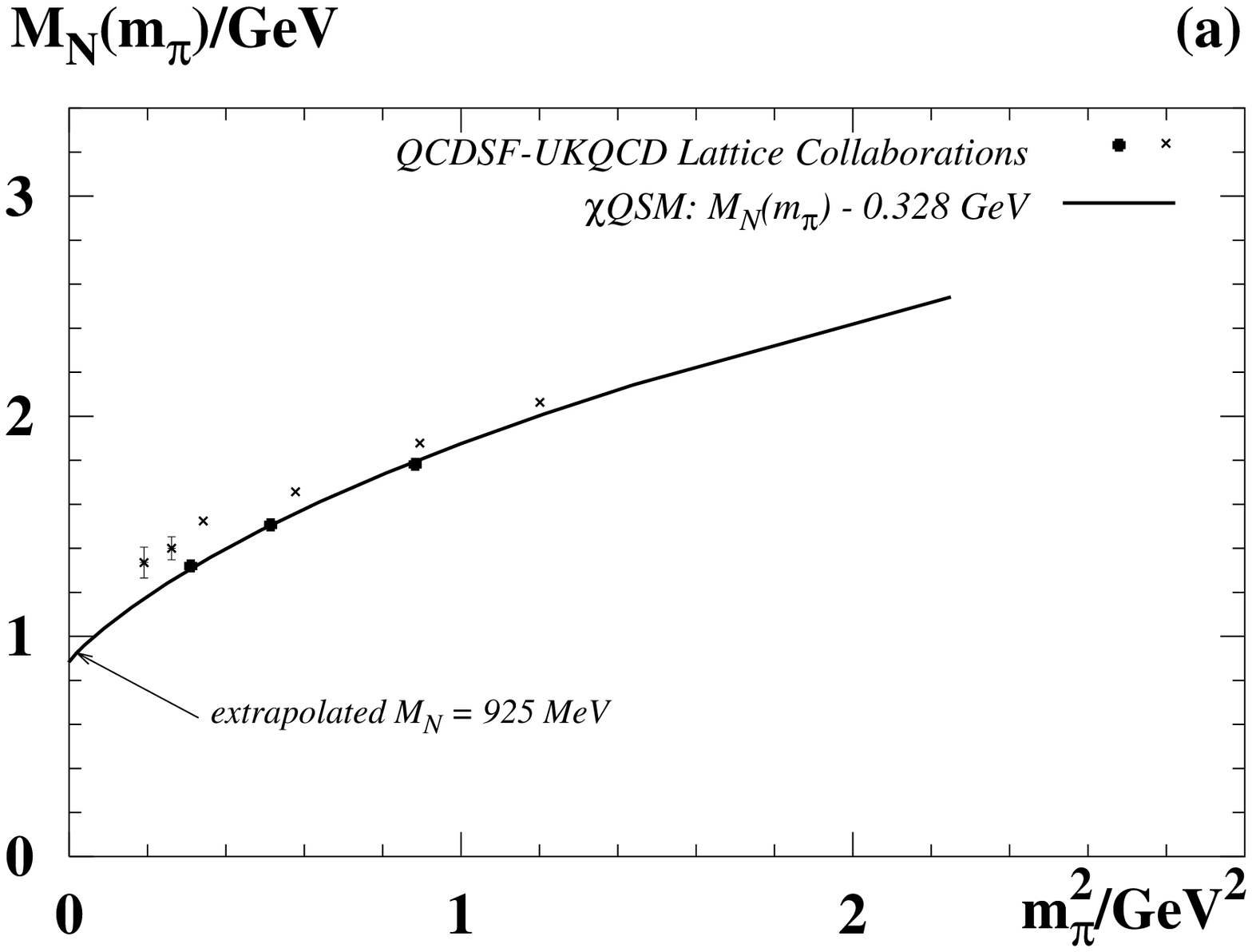}&
    \includegraphics[width=9.3cm,height=5.5cm]{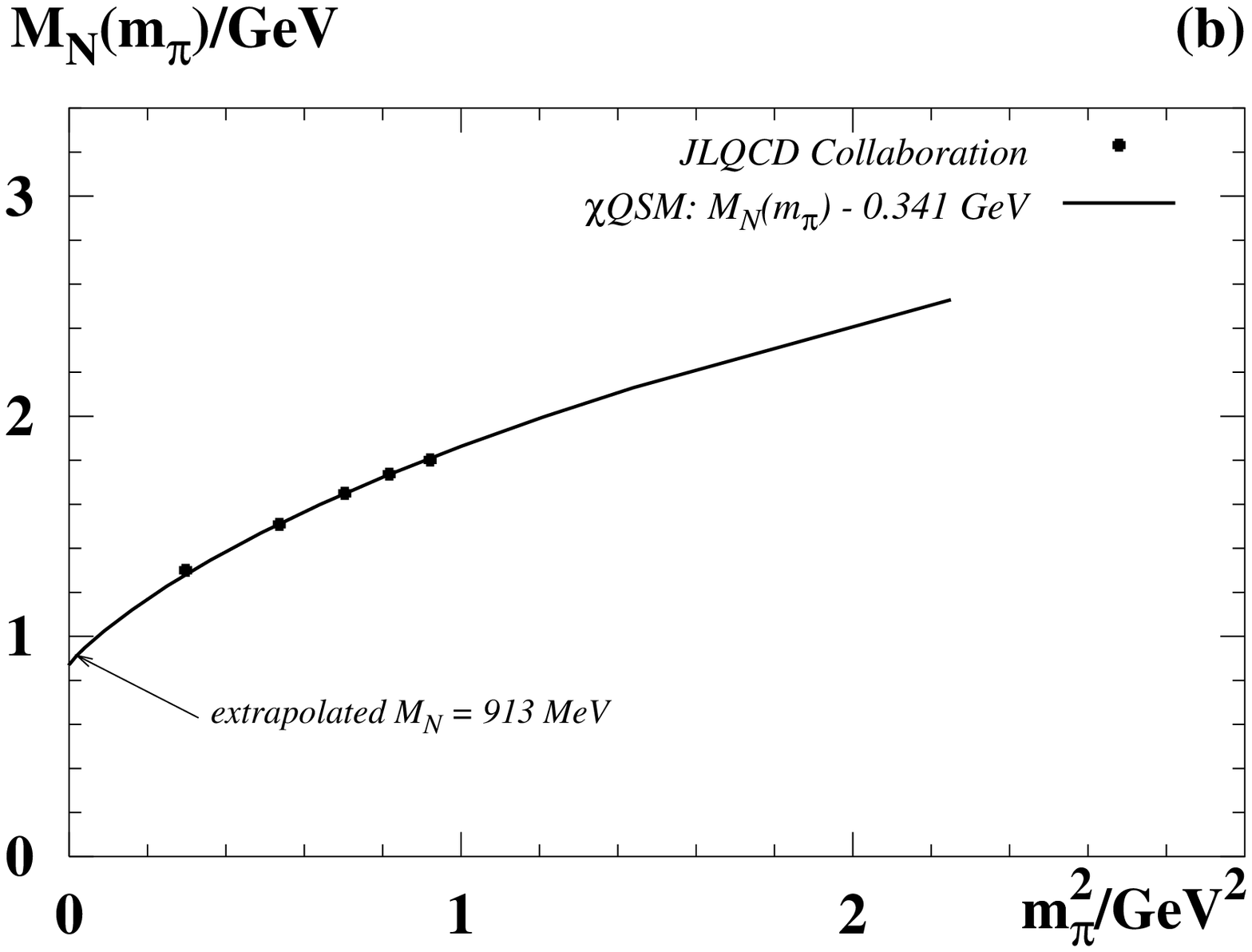}  \cr
    \includegraphics[width=9.3cm,height=5.5cm]{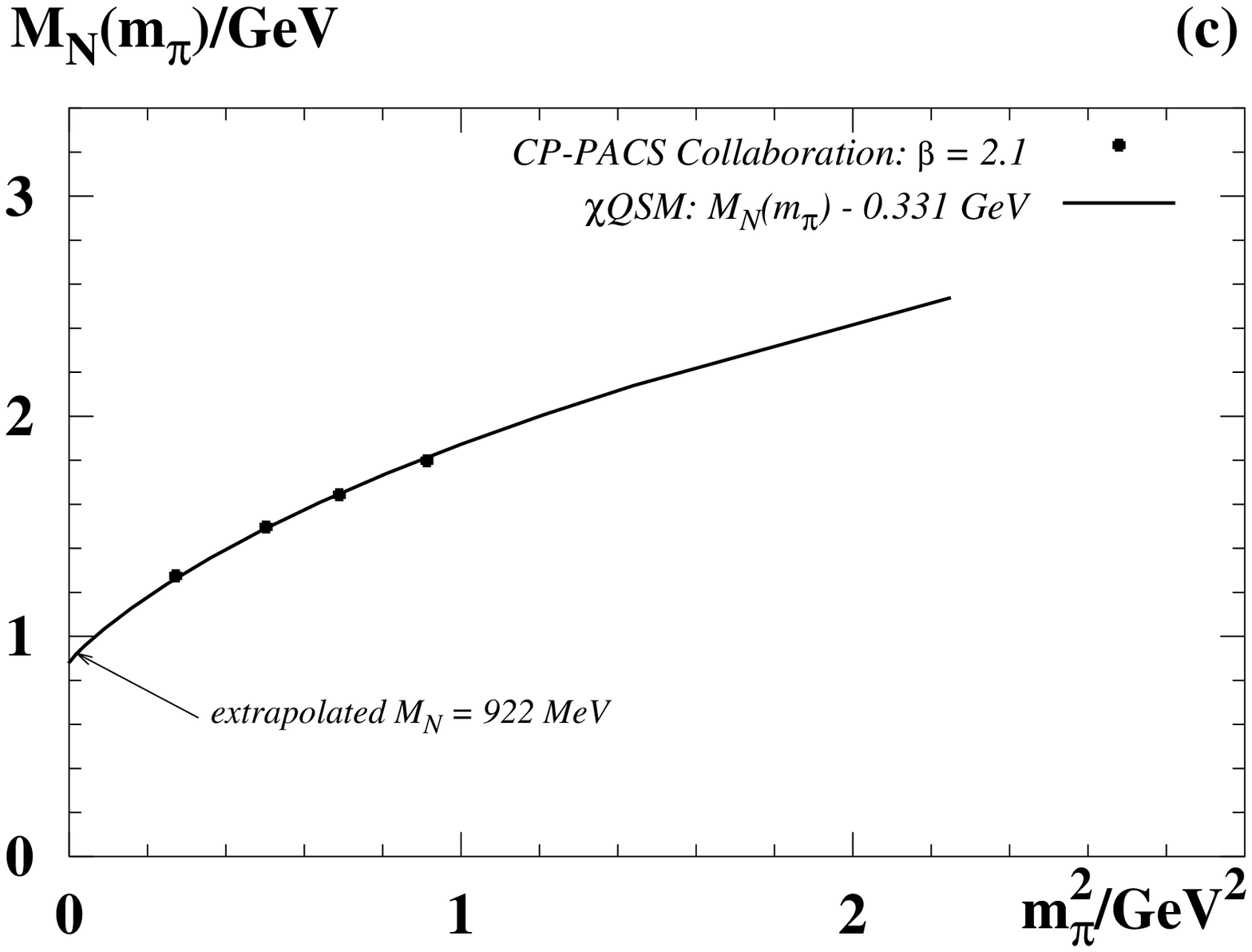}&
    \includegraphics[width=9.3cm,height=5.5cm]{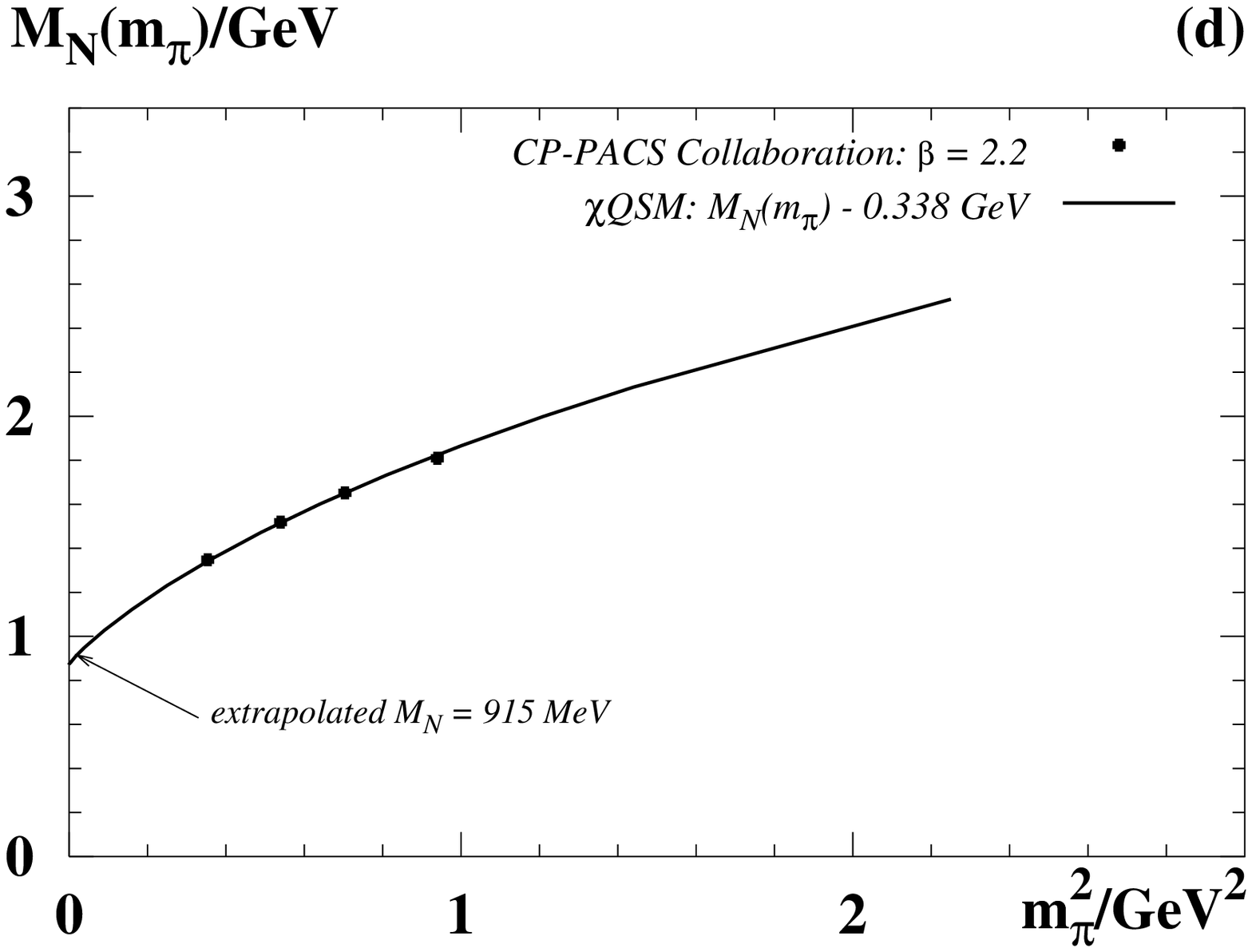}\cr
    \includegraphics[width=9.3cm,height=5.5cm]{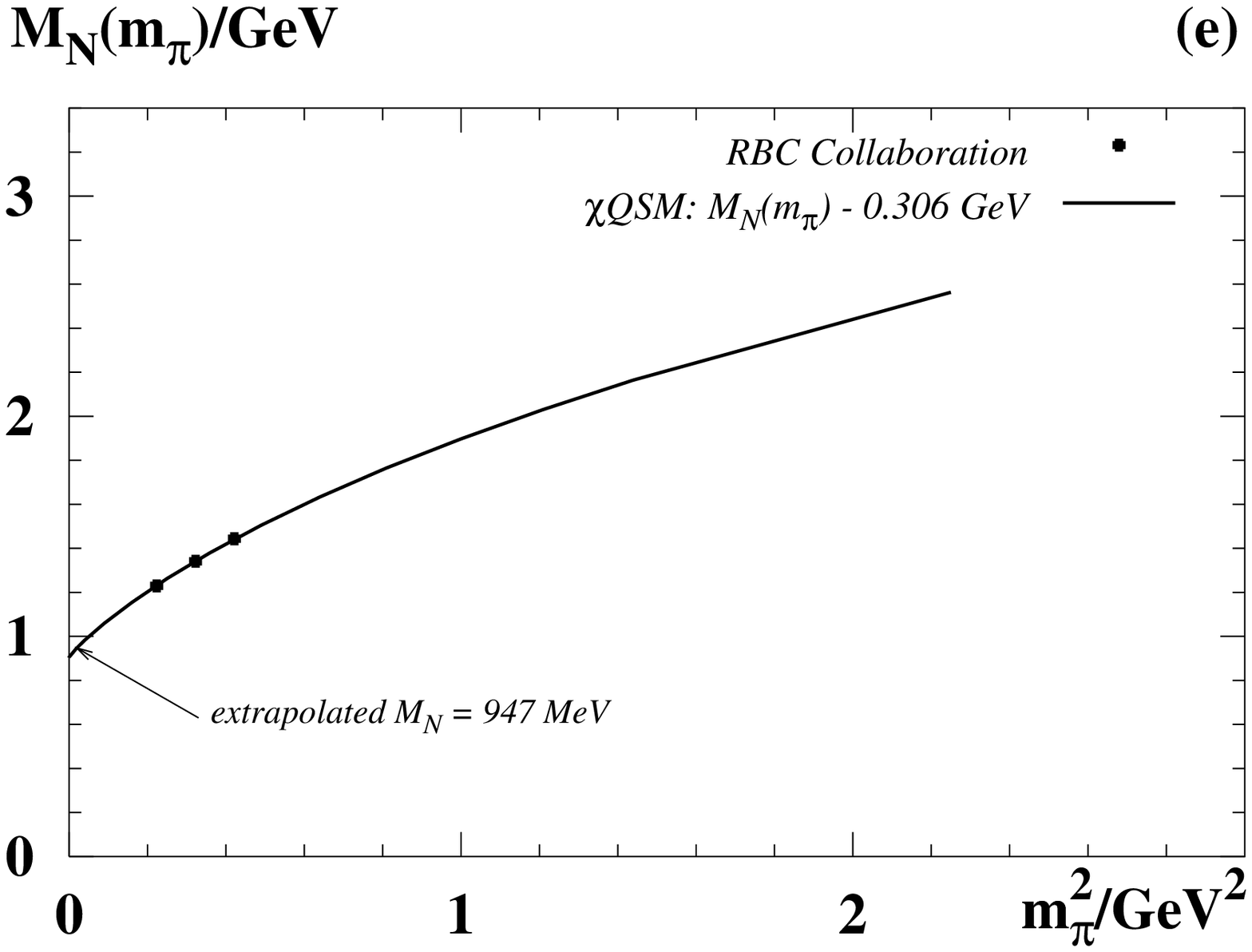}&
    \includegraphics[width=9.3cm,height=5.5cm]{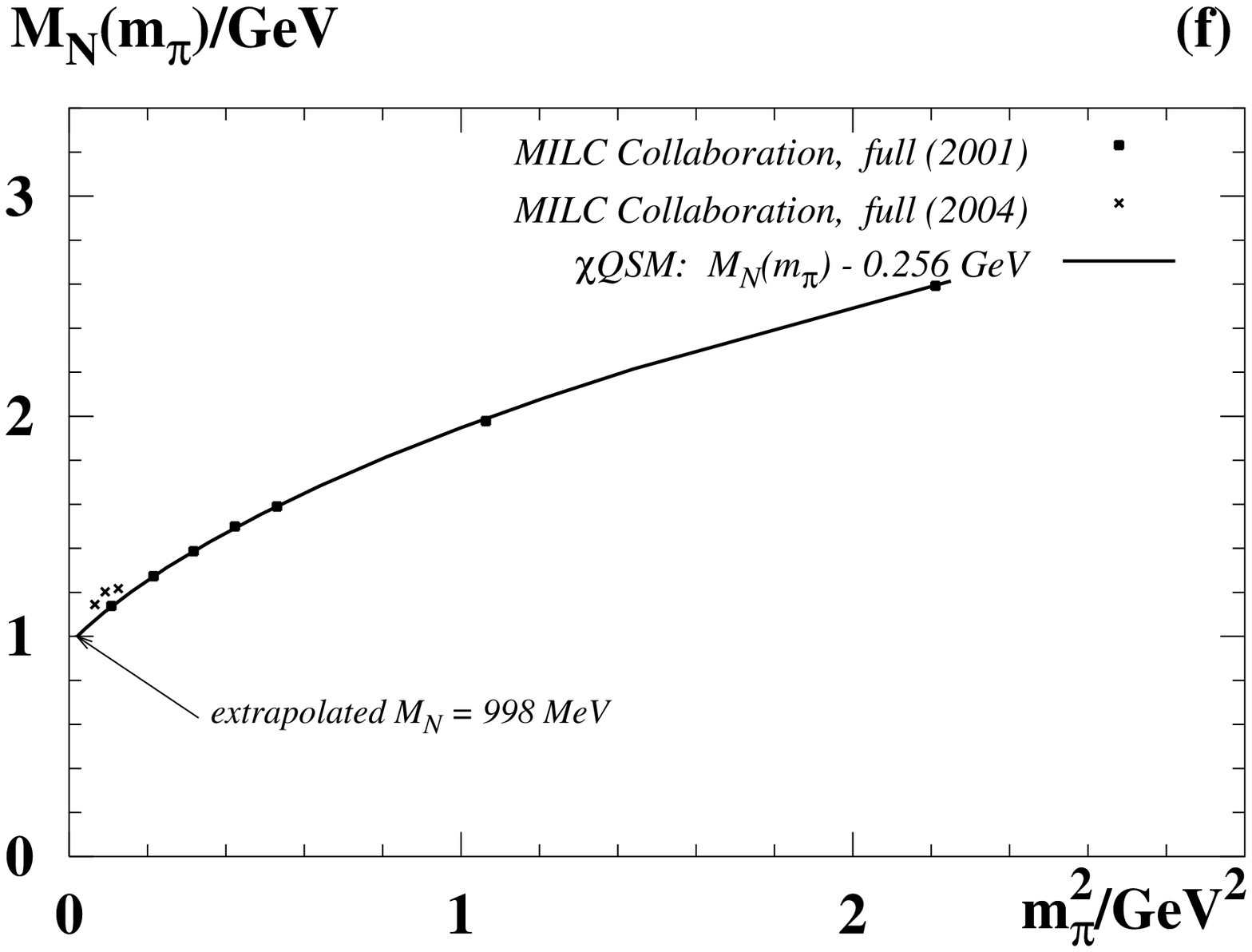}
\end{tabular}
    \caption{\label{Fig-2}
    \footnotesize\sl
    $M_N(m_\pi)$ vs.\  $m_\pi^2$ from the chiral quark-soliton model
    vs.\  lattice data on $M_N(m_\pi)$ from full simulations by
(\underline{\ref{Fig-2}a}) the UKQCD-QCDSF Collaborations
                  \cite{Allton:2001sk,AliKhan:2003cu},
(\underline{\ref{Fig-2}b}) the JLQCD Collaboration \cite{Aoki:2002uc},
(\underline{\ref{Fig-2}c},\underline{d})
                       the CP-PACS Collaboration \cite{AliKhan:2001tx},
(\underline{\ref{Fig-2}e}) the RBC Collaboration \cite{Aoki:2004ht},
(\underline{\ref{Fig-2}f}) the MILC Collaboration \cite{Bernard:2001av,Aubin:2004fs}.
All data were obtained from large lattices $L\gtrsim 2\,{\rm fm}$,
with the exception of the data marked by crosses in (\ref{Fig-2}a),
see text.
Unless error bars are shown, here and in the following figures
the statistical error of the lattice data is comparable to or
smaller than the size of the points in the plot.}
%
%
%
\begin{tabular}{cc}
    \includegraphics[width=9.3cm,height=5.5cm]{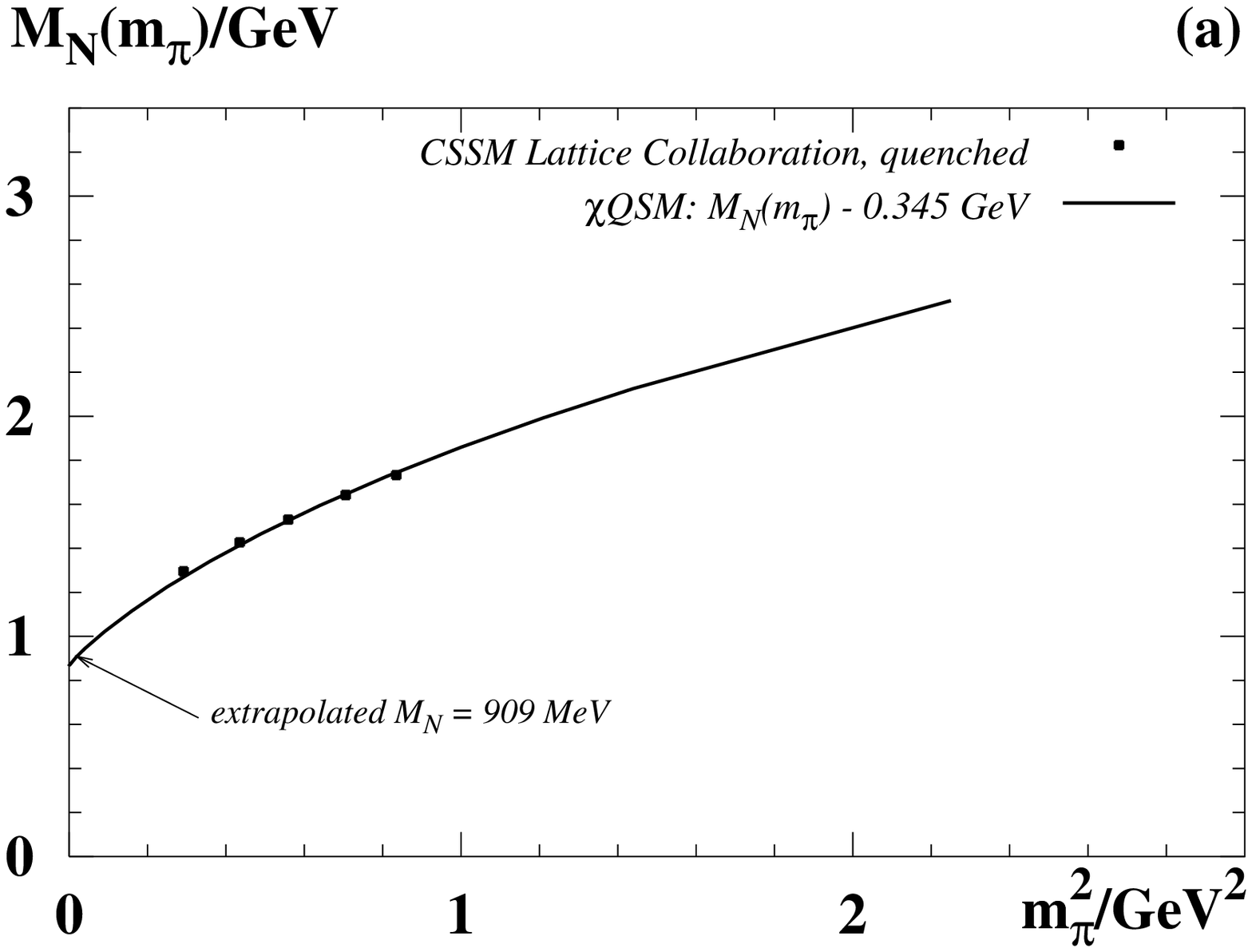}&
    \includegraphics[width=9.3cm,height=5.5cm]{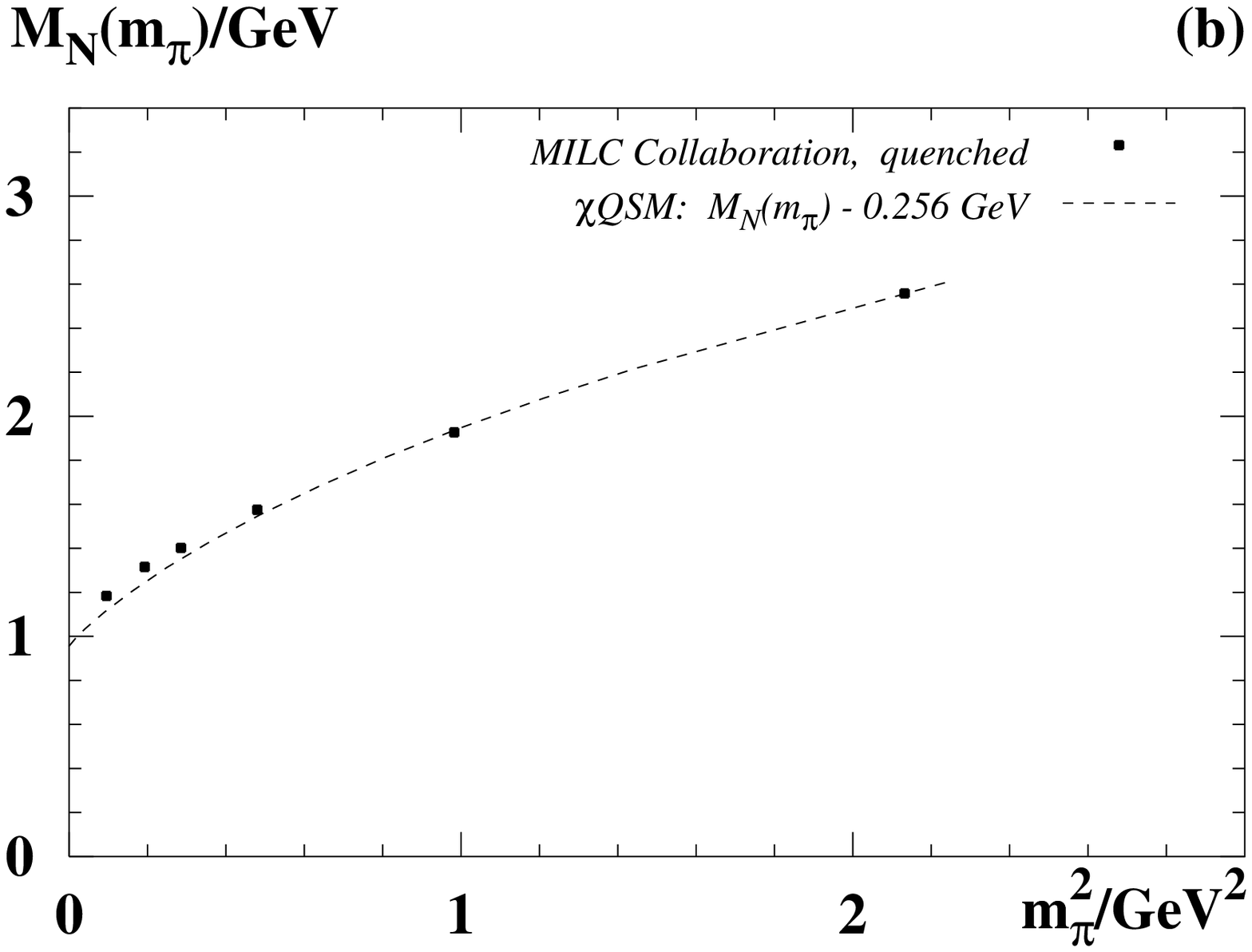}
\end{tabular}
    \caption{\label{Fig-3}
    \footnotesize\sl
    $M_N(m_\pi)$ vs.\  $m_\pi^2$ from the chiral quark-soliton model
    vs.\  lattice data on $M_N(m_\pi)$ obtained in the quenched
        approximation by
\underline{\ref{Fig-3}a} the CSSM Lattice Collaboration
                              \cite{Zanotti:2001yb} and
\underline{\ref{Fig-3}b} the MILC Collaboration \cite{Bernard:2001av}.}
\end{figure*}

$M_N(m_\pi)$ from the model overestimates the lattice data by about
$300\,{\rm MeV}$ which is expected, if we recall the discussion in
Sec.~\ref{Subsec:Mn-in-large-Nc}.
Remarkably we observe that it is possible to introduce an
$m_\pi$-{\sl independent} subtraction constant $C={\cal O}(300\,{\rm MeV})$
depending on the lattice data,
such that $[M_N(m_\pi)-C]$ agrees well with the lattice data.

%
In Fig.~\ref{Fig-2}a we compare the $\chi$QSM result for
$M_N(m_\pi)$ to full lattice data by the UKQCD and QCDSF Collaborations
\cite{Allton:2001sk,AliKhan:2003cu} obtained from simulations on
$(16-24)^3\times 48$ lattices using the standard Wilson plaquette action
for gauge fields and the non-perturbatively ${\cal O}(a)$ improved action
for fermions. The fat points in Fig.~\ref{Fig-2}a were extracted from
simulations at $5.20\le\beta\le 5.29$ on lattices of the physical size
$L\ge 1.96\,{\rm fm}$ with $m_\pi L > 6$ and lattice spacings
$a=0.09\,{\rm fm}$ to $0.12\,{\rm fm}$ fixed by the Sommer method with
$r_0=0.5\,{\rm fm}$ \cite{Sommer:1993ce}. Other lattice spacings quoted
below were also determined by means of this popular method which, however,
is not free of criticism \cite{Sommer:2003ne}.
A best fit yields $C=329\,{\rm MeV}$ with a $\chi^2$ per degree of
freedom of $\chi^2_{\rm dof} = 0.7$ and is shown in Fig.~\ref{Fig-2}a.
The uncertainty of the constant $C$ due to the statistical error
of the lattice data is of the order of magnitude of few MeV, see
Table~\ref{Table:Mn-extrapolated} and thus negligible, see the 
remarks in footnote~\ref{footnote:accuracy}.

For comparison in Fig.~\ref{Fig-2}a also data by the UKQCD and QCDSF
Collaborations \cite{Allton:2001sk,AliKhan:2003cu} are shown from smaller
lattices $L=1.5\,{\rm fm}$ to $1.7 \,{\rm fm}$. Finite size effects are
clearly visible, and were subject to a detailed study in
Ref.~\cite{AliKhan:2003cu}.
Since the present $\chi$QSM calculation by no means is able to simulate
finite volume effects, we restrict our study to data from lattices
with $L\gtrsim 2\,{\rm fm}$.
The study of Ref.~\cite{AliKhan:2003cu} indicates that for lattices
of this size finite volume effects can be assumed to be small. Also,
in this work we consider data obtained from lattices with spacings
$a\le 0.13\,{\rm fm}$ and assume discretization effects to be negligible.
Such effects are difficult to control systematically \cite{Bar:2004xp}.

%
In Fig.~\ref{Fig-2}b the $\chi$QSM result for $M_N(m_\pi)$ is compared to
lattice data by the JLQCD Collaboration from Ref.~\cite{Aoki:2002uc},
where dynamical simulations with two flavours were performed on a
$20^3\times 48$ lattice using the plaquette gauge action and the
non-perturbatively ${\cal O}(a)$ improved Wilson quark action at
$\beta = 5.20$ with lattice spacings $0.10\,{\rm fm}\le a\le 0.13\,{\rm fm}$.
The physical size of the lattice was
$1.96\,{\rm fm}\le L \le 2.53\,{\rm fm}$ and the range
$500\,{\rm MeV}\le m_\pi\le 1\,{\rm GeV}$ was covered
corresponding to $m_\pi L > 5$. The best fit yields $C=341\,{\rm MeV}$
with a $\chi^2_{\rm dof}=0.3$ and is shown in Fig.~\ref{Fig-2}b.

%
In Figs.~\ref{Fig-2}c and d we compare $M_N(m_\pi)$ from the model to lattice
results by the CP-PACS Collaboration \cite{AliKhan:2001tx}, which were
obtained on $24^3\times 48$ lattices from a renormalization-group improved
gauge action and a mean field improved clover quark action at $\beta=2.1$ (and
$\beta = 2.2$) with dynamical $u$- and $d$-quarks and quenched $s$-quarks.
For hadrons with no valence $s$-quarks this practically means a full
two-flavour simulation. The physical lattice spacings and sizes
were $a = 0.09\,{\rm fm}$ to $0.13\,{\rm fm}$ and
$L = 2.22\,{\rm fm}$ to $3.12\,{\rm fm}$ with pion masses in the range
$500\,{\rm MeV} \le m_\pi \le 1\,{\rm GeV}$, such that $m_\pi L \gtrsim 7$.
We observe that $M_N(m_\pi)$ from the $\chi$QSM with a subtraction
constant $C=331\,{\rm MeV}$ with $\chi^2_{\rm dof}=2.2$
(and     $C=338\,{\rm MeV}$ with $\chi^2_{\rm dof}=0.4$)
describes well the lattice data of Ref.~\cite{AliKhan:2001tx},
see Fig.~\ref{Fig-2}c (and Fig.~\ref{Fig-2}d).

%
Next we confront the model results for $M_N(m_\pi)$ to lattice data
from dynamical two-flavour simulations with domain wall fermions by the
RBC Collaboration \cite{Aoki:2004ht}, which have the virtue of preserving
chiral invariance. In Ref.~\cite{Aoki:2004ht} a renormalization group
improved (``doubly blocked Wilson'') gauge action with $\beta=0.80$ was used
on a $16^3\times 32$ lattice with the physical spacing of $0.12\,{\rm fm}$
and a lattice size about $2\,{\rm fm}$. The range of pion masses was
$m_\pi=470\,{\rm MeV}$ to $650\,{\rm MeV}$. A best fit yields for the
constant $C=306\,{\rm MeV}$ with a $\chi^2_{\rm dof}=0.02$, and provides
a very good description of the data, see Fig.~\ref{Fig-2}e.

%
In Fig.~\ref{Fig-2}f we confront $M_N(m_\pi)$ from the $\chi$QSM
with 2001 lattice data by the MILC Collaboration \cite{Bernard:2001av}
obtained from simulations with three dynamical quarks using a one-loop
Symanzik improved gauge action and an improved Kogut-Susskind quark
action. The physical lattice size was tuned to $L=2.6\,{\rm fm}$ with
a lattice spacing of $0.13\,{\rm fm}$ and the range of pion masses
$340\,{\rm MeV} \le m_\pi \le 2.2\,{\rm GeV}$ was covered.
The best fit yields $C=256\,{\rm MeV}$ for the subtraction constant
(with $\chi^2_{\rm dof} = 1.2$).

One could worry whether the SU(2) model results can be compared to
three-flavour lattice simulations, though in \cite{Bernard:2001av}
for the nucleon mass no significant differences were noticed between
two- and three-flavour runs.
However, as noted in Sec.~\ref{Sec:2-pion-nucleon-in-eff-theory}, the
nucleon mass is the same in the SU(2) and SU(3) versions of the $\chi$QSM if
$m_u=m_d=m_s$, which is the case for the lattice data \cite{Bernard:2001av}
for $m_\pi > 700\,{\rm MeV}$. Restricting the fitting procedure to this
range of $m_\pi$ (the last three points in Fig.~\ref{Fig-2}f) we obtain
$C=260\,{\rm MeV}$ (with a $\chi^2_{\rm dof} = 1.9$). This fit is shown
in Fig.~\ref{Fig-2}f. Note that it is practically indistinguishable, c.f.\
footnote~\ref{footnote:accuracy}, from the fit where the whole $m_\pi$-range
(i.e.\  also data with $m_u=m_d<m_s$) was used, and equally well describes
the region of lower $m_\pi$. In any case we observe a good agreement with
the lattice data \cite{Bernard:2001av} up to $m_\pi = 1.5\,{\rm GeV}$.

We include in Fig.~\ref{Fig-2}f also the recent small-$m_\pi$ (2004-)
MILC data \cite{Aubin:2004fs} from the ``coarse'' lattices with $L=20$
(or $L=24$ for the lowest $m_\pi$-value) which have a lattice spacing
$a\approx 0.12\,{\rm fm}$ comparable to \cite{Bernard:2001av}.  These
data are not compatible with the $\chi$QSM result.  In this context it
is important to note that the simulation for the highest $m_\pi$-value
of the 2004 MILC data \cite{Aubin:2004fs} is an extended run of the
simulation for the lowest $m_\pi$ of the 2001 data
\cite{Bernard:2001av}.  One would therefore expect that they coincide
in the plot in Fig.~\ref{Fig-2}f, which is not the case.  In fact, the
MILC data \cite{Bernard:2001av,Aubin:2004fs} for this particular
simulation are well consistent with each other in {\sl lattice units}:
$am_\pi$ and $aM_N$ from these runs agree within statistical error bars.

The discrepancy of the results from \cite{Bernard:2001av,Aubin:2004fs} 
in Fig.~\ref{Fig-2}f is due to the different values for $r_1$ --- 
a parameter defined similarly to the Sommer scale $r_0$ and used 
to fix the physical units in \cite{Bernard:2001av,Aubin:2004fs}.
Different methods were used to determine the physical unit of $r_1$
--- resulting in the value $r_1=0.35\,{\rm fm}$ in \cite{Bernard:2001av}, 
and $r_1=0.324(4)\,{\rm fm}$ in \cite{Aubin:2004fs}.
Had we used the $r_1$-value from \cite{Bernard:2001av} to give physical 
units to the dimensionless lattice numbers for $am_\pi$ and $aM_N$ 
from \cite{Aubin:2004fs}, then the two data sets would perfectly agree
in Fig.~\ref{Fig-2}f, and the 2004 data would be compatible with the
$\chi$QSM results. The precise determination of the physical units of 
lattice data is a difficult issue \cite{Sommer:1993ce,Sommer:2003ne}, 
see also \cite{Bernard:2001av,Aubin:2004fs}.

%
\begingroup
\squeezetable
\begin{table*}[t!]
    	\caption{\footnotesize\sl
   	\label{Table:Mn-extrapolated}
   Comparison of the $M_N(m_\pi)$ obtained from the $\chi$QSM to the
   lattice data \cite{AliKhan:2001tx,Bernard:2001av,Aoki:2002uc,Allton:2001sk,Aoki:2004ht,AliKhan:2003cu,Zanotti:2001yb}.
   For convenience we quote the lattice sizes and spacings and the range of 
   $m_\pi$ covered in the lattice simulations in physical units which
   were fixed by the Sommer method \cite{Sommer:1993ce}. 
   The soliton approach generally overestimates \cite{Pobylitsa:1992bk}
   the nucleon mass at the physical point by about $300\,{\rm MeV}$, see 
   Eq.~(\ref{Mn-at-mpi=140MeV}). We find a similar overestimate at the 
   respective lattice values of $m_\pi$. Correcting for this overestimate
   by introducing a $m_\pi$-independent subtraction constant $C$ to be
   fitted to the respective lattice data set, we observe a good agreement 
   $[M_N(m_\pi)-C]$ with the lattice data, see Figs.~\ref{Fig-2} and 
   \ref{Fig-3}. 
   The 5th row shows the fit results for the constant $C$ and its 1-$\sigma$ 
   uncertainty due to the statistical error of the lattice data, and the 6th 
   row shows the $\chi^2$ per degree of freedom ($\chi^2_{\rm dof}$) of the 
   respective fit.
   Also the ``extrapolated'' (within the $\chi$QSM) value of the nucleon 
   mass at the physical point ($M_N$) is included. It has the same 
   uncertainty as the fit-constant $C$, which is due to the statistical 
   error of lattice data and practically negligible, see the remark in 
   footnote~\ref{footnote:accuracy}.  
   It has also an unestimated systematic error due to model-dependence, 
   see Sec.~\ref{Sec:4-using-for-extrapolation?}.}
\vspace{0.2cm}
    \begin{ruledtabular}
    \begin{tabular}{lccclcc}
    \\
    Collaboration \hspace{-0.7cm} &
    lattice size/fm	&
    spacing/fm 		&
    $m_\pi$-range/MeV	& 
    $C$/MeV & 
    $\chi^2_{\rm dof}$ &
    $M_N$/MeV \\
          &&&&&\\
    \hline&&&&&\\
UKQCDSF \cite{Allton:2001sk,AliKhan:2003cu} & 2.0-2.2 & 0.09-0.12 & 550-940  & $329\pm 7$  & 0.7 & 925 \\
JLQCD   \cite{Aoki:2002uc}                  & 2.0-2.5 & 0.10-0.13 & 540-960  & $341\pm 7$  & 0.3 & 913 \\
CP-PACS \cite{AliKhan:2001tx} ($\beta=2.1$) & 2.7-3.1 & 0.11-0.13 & 520-960  & $331\pm 4$  & 2.2 & 922 \\
CP-PACS \cite{AliKhan:2001tx} ($\beta=2.2$) & 2.2-2.4 & 0.09-0.10 & 590-970  & $338\pm 6$  & 0.4 & 915 \\
RBC     \cite{Aoki:2004ht}                  & 1.9-2.0 &    0.12   & 470-650  & $306\pm 11$ & 0.02& 947 \\
MILC \cite{Bernard:2001av} (all data)       & 2.6     &    0.13   & 350-1500 & $256\pm 2$  & 1.2 & 998 \\
MILC \cite{Bernard:2001av} (only $m_{u,d}=m_s$) & 2.6 &    0.13   & 720-1500 & $260\pm 3$  & 1.9 & 993 \\
CSSM \cite{Zanotti:2001yb} (quenched)       & 2.0     &    0.125  & 540-920  & $345\pm 5$  & 1.3 & 909 \\
         \\
\end{tabular}
\end{ruledtabular}
\end{table*}
\endgroup
%

It is instructive to compare the model results also to lattice data obtained
from simulations performed in the {\sl quenched} approximation, e.g.,
by the CSSM Lattice Collaboration \cite{Zanotti:2001yb}, where
a mean-field improved gauge action and a fat-link clover fermion (``FLIC'')
action was used on a $16^3\times 32$ lattice with a lattice spacing
of $a=0.125\,{\rm fm}$. The calculation covers the range
$540\,{\rm MeV}\le m_\pi\le 920\,{\rm MeV}$.
A best fit to the quenched data gives $C=345\,{\rm MeV}$
(with a $\chi^2_{\rm dof}=1.3$) which yields a good
agreement with the lattice data, see Fig.~\ref{Fig-3}a.

For the quenched data by the MILC Collaboration \cite{Bernard:2001av},
however, we observe that a fit would work much worse. In this case we
refrain from fitting and show instead in Fig.~\ref{Fig-3}b the fit to
the full MILC data from Fig.~\ref{Fig-2}f, which nicely illustrates
how results from full and quenched calculations differ.
Interestingly, at large $m_\pi^2$ the full and quenched data
of Ref.~\cite{Bernard:2001av} agree well with eachother.
In fact, it is not suprising that differences between full and quenched
simulations become less pronounced with increasing $m_\pi^2$, i.e.\
with increasing fermion masses.

Thus, we observe that the $\chi$QSM-result for $M_N(m_\pi)$ 
supplemented by an $m_\pi$-independent subtraction constant $C$ 
(whose precise value follows from a best fit to the respective lattice data) 
is able to describe the lattice data \cite{AliKhan:2001tx,Bernard:2001av,Aoki:2002uc,Allton:2001sk,Aoki:2004ht,AliKhan:2003cu,Zanotti:2001yb} 
over a wide range of pion masses $350\,{\rm MeV}\le m_\pi\le 1500\,{\rm MeV}$.
The results are shown in Figs.~\ref{Fig-2} and \ref{Fig-3}
and are summarized in Table~\ref{Table:Mn-extrapolated}.

\section{\boldmath $\chi$QSM as tool for extrapolation?}
\label{Sec:4-using-for-extrapolation?}

From the $[M_N(m_\pi)-C]$ at $m_\pi=140\,{\rm MeV}$ with the constant $C$
fitted to the respective data, we can read off in principle the physical value 
of the nucleon mass -- ``extrapolated'' from the resüpective lattice data
by means of the $\chi$QSM as a guideline. These extrapolated values are 
indicated by arrows in 
Figs.~\ref{Fig-2}a-f and \ref{Fig-3}a, and are summarized in 
Table~\ref{Table:Mn-extrapolated}.

It is  worthwhile stressing that the extrapolated values in
Table~\ref{Table:Mn-extrapolated} agree within an accuracy of $\pm 5\%$
with the physical nucleon mass. 
(Though at the same time, lattice data for $f_\pi$ are underestimated by 
up to $40\%$ at large $m_\pi\sim 1\,{\rm GeV}$, see Sec.~\ref{Subsec-fix-para}.)
The uncertainty of the extrapolated values for the nucleon mass is the same 
as for the fit constant $C$ as quoted in Table~\ref{Table:Mn-extrapolated},
i.e.\  it is of the order of few MeV and thus negligible, 
see footnote~\ref{footnote:accuracy}.  

However, what does not need to be negligible, is the systematic error of 
such an extrapolation. First of all, this extrapolation is done within the 
$\chi$QSM, and the results are model-dependent. 
E.g. handling the model parameters in the chiral limit differently would 
change our results --- though one may expect a qualitatively similar picture,
see the discussion in Sec.~\ref{Subsec-fix-para}. 
Apart from this source of model-dependence, which will be subject to
future numerical studies \cite{work-in-progress}, there are principal 
difficulties to estimate reliably the systematic error within the model, 
as we shall see in the following.

Let us first address the question of the range of reliability of the 
$\chi$QSM-description of $M_N(m_\pi)$. For that it is instructive
to compare to $\chi$PT and the effective FRR approach.
$\chi$PT to ${\cal O}(p^4)$ was argued to provide a 
reliable expansion for $M_N(m_\pi)$ up to $m_\pi^2 < 0.4\,{\rm GeV}$
\cite{Bernard:2003rp,Procura:2003ig,Bernard:2005fy,Frink:2004ic} 
($p$ is to be identified with the generic small expansion parameter 
in $\chi$PT, e.g., in our context pion mass). 
A more conservative bound $m_\pi^2 < 0.1\,{\rm GeV}^2$ was given 
in \cite{Beane:2004ks}. The effective FRR approach was argued to 
correspond to a {\sl partial} resummation of the chiral expansion, and 
to be valid up to $m_\pi^2 < 1\,{\rm GeV}^2$ \cite{Leinweber:2003dg}
(which comes, of course, at the prize of introducing model dependence,
see the discussion in Sec.~\ref{Sec:1-introduction}.)

\begin{figure}[b!]
\vspace{-0.5cm}
\includegraphics[width=5cm,height=5cm]{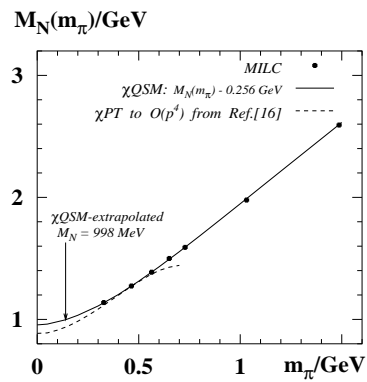}
    \caption{\footnotesize\sl
    \label{Fig-NEW} 
    Lattice data by MILC \cite{Bernard:2001av} on $M_N(m_\pi)$ vs.\ $m_\pi$ 
    and the fits to these data in $\chi$PT from Ref.~\cite{Bernard:2005fy}
    (where the physical value of $M_N$ was used as input)
    and in the  $\chi$QSM, c.f.\  Fig.~\ref{Fig-2}f.
    The figure illustrates the wide range of applicability of the $\chi$QSM.}
\end{figure}

What might be the range of reliability of the $\chi$QSM?
Recall that the $\chi$QSM expression for the nucleon mass
(as well as for any other quantity) may be considered as a resummed
infinite series in derivatives of the chiral field $U=\exp(i\tau^a\pi^a)$,
c.f.\  Eq.~(\ref{Eq:gradient-expansion}), whereby it is understood that each
chiral order is evaluated in the large-$N_c$ limit. (Keep in mind 
that these limits do not commute, see Sect.~\ref{Subsec:chiral-limit}.)
Thus, one may argue that the $\chi$QSM corresponds to a chiral expansion,
which is {\sl completely} resummed  -- to leading order of the
large-$N_c$ expansion. If one were happy with this approximation, then
$\chi$QSM results could be considered reliable for all $m_\pi$ including
the heavy quark limit, see the discussion in Sec.~\ref{Subsec:heavy-quark}.
It must be stressed, that -- as in the case of the FRR approach  
-- this chiral resummation is performed {\sl within} a particular  
model of the nucleon. Thus, our results and conclusions are 
inevitably model dependent.

The wide range of reliability of the $\chi$QSM we observe in practice is illustrated 
in Fig.~\ref{Fig-NEW}, where we show the $\chi$QSM-fit to the MILC lattice data 
\cite{Bernard:2001av} and the $\chi$PT-fit from \cite{Bernard:2005fy}.
(For further details see Sec.~\ref{Sec:5-compare-to-CPT}.)

The perhaps most serious restriction might be that the $\chi$QSM assumes
the number of colours $N_c$ to be large. What might be the effects due 
to $1/N_c$-corrections?

Let us consider the mass difference between the $\Delta$-resonance
and the nucleon as a measure for such corrections. Note that this
mass difference vanishes not only in the large-$N_c$ limit,
see Eq.~(\ref{Eq:Mdelta-Mn}), but also in the heavy quark limit
\cite{Dashen:1993jt}. This is supported by lattice results
\cite{Bernard:2001av,AliKhan:2001tx,Zanotti:2001yb}, where always
$\Delta \equiv M_\Delta-M_N < m_\pi$ holds.
(This means that on present day lattices the $\Delta$-resonance
is safe from strong decays and thus a stable particle.)
In the region $m_\pi^2 > 0.32 \,{\rm GeV}^2$ one finds $\Delta < \frac15m_\pi$
\cite{Bernard:2001av,AliKhan:2001tx,Zanotti:2001yb}.
Thus, practically for most of the present day lattice data the condition
\be\label{Eq:Mdelta-Mn-on-lattice}
    \Delta\equiv M_\Delta-M_N \ll m_\pi \;\;\mbox{for}\;\; 
    m_\pi^2 > 0.32\,{\rm GeV}^2
\ee
is satisfied, such that the $\Delta$-nucleon mass difference can be
neglected to a good approximation. This may be a reason for the good
description of lattice data in the $\chi$QSM up to
$m_\pi^2\le 2.3\,{\rm GeV}^2$ in Sec.~\ref{Sec:3-compare-to-lattice}.

However, in the physical region $\Delta =290\,{\rm MeV}$ is larger than
$m_\pi=140 \,{\rm MeV}$, and the leading order large-$N_c$ treatment of the
nucleon mass inevitably introduces a serious systematic error. In fact, 
one could express the nucleon mass as a function of $m_\pi$ 
and $y=\Delta/m_\pi$ as
\be\label{Eq:Mn-as-funct-of-mpi-Delta}
    M_N(m_\pi,\Delta) = F(m_\pi,y) \;,\;\;\; y=\frac{\Delta}{m_\pi}\;.
\ee
Then, using the $\chi$QSM as a guideline for the extrapolation of
lattice data corresponds to approximating
\be\label{Eq:Mn-as-funct-of-mpi-Delta-2}
    M_N = F(m_\pi,2.1) \approx  F(m_\pi,0) \;\;\mbox{at}\;\;
    m_\pi=140\,{\rm MeV}.
\ee
It is difficult to quantify the systematic error associated with this
approximation. A very rough estimate of this error within the model 
is given in Appendix~\ref{App-C}.

Finally, let us discuss the role of the subtraction constant $C$.
As mentioned in Sec.~\ref{Sec:2-pion-nucleon-in-eff-theory} the appearence 
of such a constant is theoretically well motivated and understood in the
soliton approach \cite{Pobylitsa:1992bk}.
The comparison to lattice data indicates that this constant is about
$C={\cal O}(300\,{\rm MeV})$ and $m_\pi$-independent in the covered 
$m_\pi$-range within the statistical accuracy of the lattice data, 
see Table~\ref{Table:Mn-extrapolated}.
Although this happens to be the magnitude for this constant needed for the 
model result to coincide with the physical value of the nucleon mass, see 
Eq.~(\ref{Mn-at-mpi=140MeV}), it would be premature to assume the constant 
$C$ to be $m_\pi$-independent for all $m_\pi$. However, this is what we
implicitly we did when quoting
the extrapolated values for $M_N$ in Table~\ref{Table:Mn-extrapolated}.
The question, whether or not the constant $C$ is $m_\pi$ dependent,
cannot be answered rigorously within the model.

As an intermediate summary, we conclude that besides the general drawback
of being model-dependent, the use of the $\chi$QSM as a guideline for
an extrapolation of lattice data is limited by two major sources
of systematic error, namely $1/N_c$ corrections and a possible 
$m_\pi$-dependence of the constant $C$. Both are not under control 
within the model and prevent a reliable estimate of the systematic 
error of the extrapolation. At this point it is instructive to compare 
to $\chi$PT and the FRR model, which may give us a rough idea 
about the size of the systematic effects.

\section{Comparison to \boldmath $\chi$PT and FRR}
\label{Sec:5-compare-to-CPT}

In order to test the results from the $\chi$QSM below the $m_\pi$ range available from 
lattice QCD we have to compare to $\chi$PT which allows to connect model-independently
lattice data through the physical point to the chiral limit. In the following 
we shall focus on the analyses \cite{AliKhan:2003cu,Procura:2003ig,Bernard:2005fy} 
in $\chi$PT up to ${\cal O}(p^4)$ (and refer to them simply as $\chi$PT).
To this order the chiral expansion of the nucleon mass is characterized by 4 
low-energy constants --- or more precisely, by 4 linearly independent combinations 
of them. These constants are in principle known from studies of nucleon-nucleon or 
pion-nucleon low-energy scattering data. However, they can alternatively be determined
from a fit to lattice data.

\begin{figure*}[t!]
\begin{tabular}{rr}
\hspace{-0.5cm}\includegraphics[width=9.3cm,height=6.9cm]{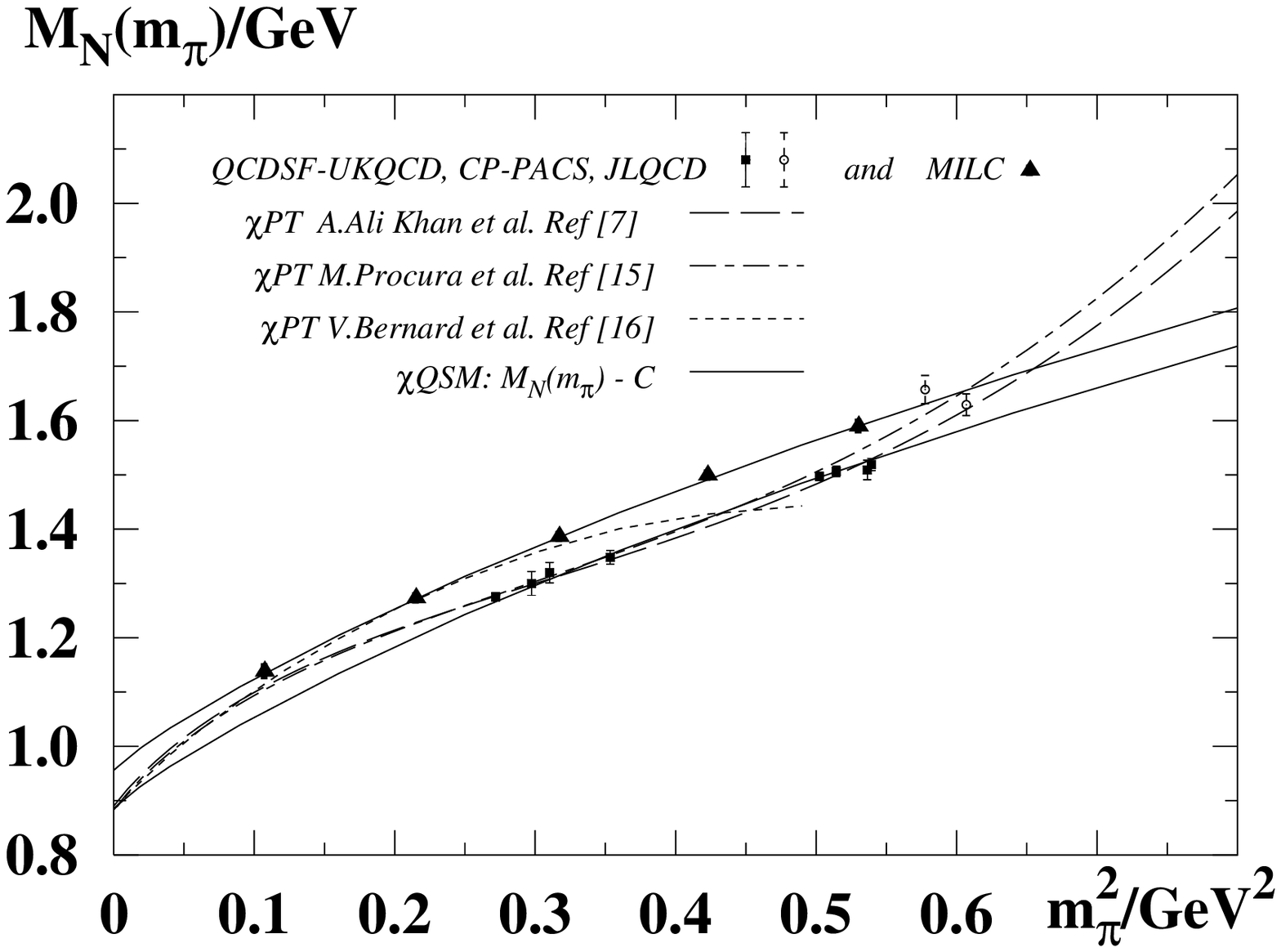}&
\hspace{-1cm}\includegraphics[width=10.5cm,height=6.9cm]{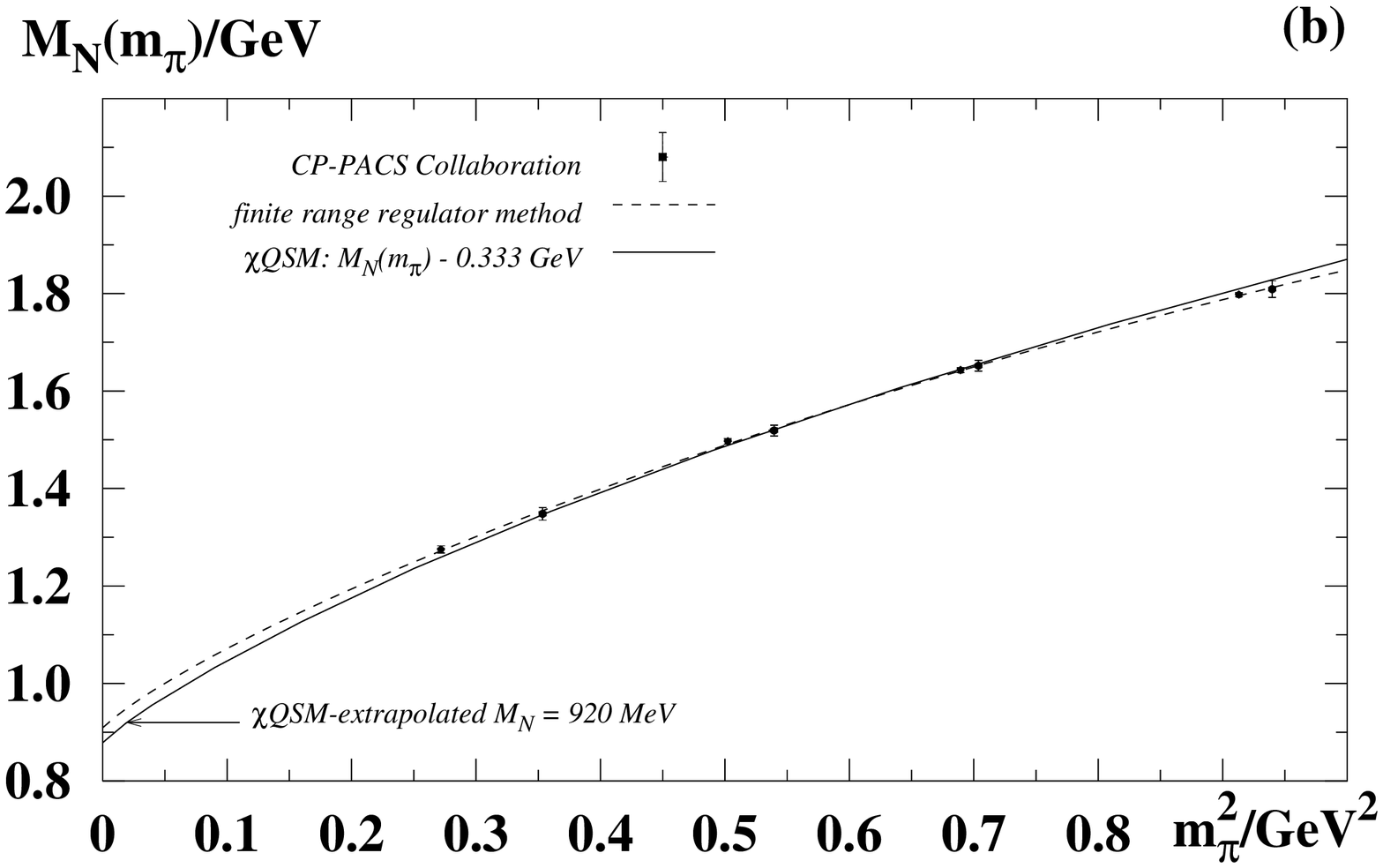}
\end{tabular}
  \caption{
  \label{Fig-4}
  \footnotesize\sl
  (a) Selected lattice data on $M_N(m_\pi)$ by QCDSF-UKQCD, CP-PACS, JLQCD 
  \cite{AliKhan:2001tx,Aoki:2002uc,Allton:2001sk,AliKhan:2003cu}
  and the corresponding fits in $\chi$PT to ${\cal O}(p^4)$ from 
  Refs.~\cite{AliKhan:2003cu,Procura:2003ig},
  in comparison lattice data by MILC \cite{Bernard:2001av}
  and the fit in $\chi$PT to ${\cal O}(p^4)$ from Ref.~\cite{Bernard:2005fy}.
  (b) Lattice data by CP-PACS \cite{AliKhan:2001tx} and the fit
  in the finite range regulator approach (with a dipole regulator)
  \cite{Leinweber:2003dg}.
  In both figures the respective $\chi$QSM-fits are shown.}
\end{figure*}

Fig.~\ref{Fig-4}a shows the $M_N(m_\pi)$ as obtained in $\chi$PT from fits to the 
lattice data \cite{AliKhan:2001tx,Aoki:2002uc,Allton:2001sk,AliKhan:2003cu} satisfying
$a < 0.15\,{\rm fm}$, $m_\pi L > 5$ and respectively the constraint
$m_\pi<800\,{\rm MeV}$ in Ref.~\cite{AliKhan:2003cu}, and
$m_\pi<600\,{\rm MeV}$ in Ref.~\cite{Procura:2003ig}.
Also shown in Fig.~\ref{Fig-4}a is the fit of Ref.~\cite{Bernard:2005fy}
to the MILC data \cite{Bernard:2001av}, where a simultaneous fit to lattice 
data on $M_\Delta(m_\pi)$ was performed.
In \cite{Procura:2003ig,Bernard:2005fy} the physical values of the nucleon 
and/or $\Delta$-mass were included as constraints to the fit.
%
Note that low-energy constants resulting from the different fits 
\cite{AliKhan:2003cu,Procura:2003ig,Bernard:2005fy} are compatible with 
the respective phenomenological values. From this point of view the 
different lattice data in Fig.~\ref{Fig-4}a are consistent with each other.
For comparison in Fig.~\ref{Fig-4}a our $\chi$QSM-fits to the same data sets are 
shown.

Notice that the two highest-$m_\pi$ data points
(marked by empty circles in Fig.~\ref{Fig-4}a) by the UKQCD
Collaboration \cite{Allton:2001sk,AliKhan:2003cu} were obtained from somehow
smaller lattices of the size $L=1.56\,{\rm fm}$ and $1.68\,{\rm fm}$ and
clearly do not follow the tendency of the $M_N(m_\pi)$ from the $\chi$QSM.
These points are therefore omitted from the fit shown in Fig.~\ref{Fig-4}a.
Including these points significantly worsens the $\chi^2_{\rm dof}$ of the
fit, see Table~\ref{Table:CQSM-vs-CPT} where the results are summarized.
Since the $\chi$QSM-description of the lattice data effectively works
also at significantly larger $m_\pi$, see Secs.~\ref{Sec:3-compare-to-lattice}
and \ref{Sec:4-using-for-extrapolation?} and c.f.\  Fig.~\ref{Fig-NEW}, 
we conclude that, quoting Ref.~\cite{Procura:2003ig},
``{\sl the surprizingly good (and not yet understood) agreement with
lattice data} [above $m_\pi>600\,{\rm MeV}$] {\sl even up to 
$m_\pi\approx 750$ MeV}'' is an accidental consequence of comparing 
$\chi$PT to ${\cal O}(p^4)$ at the edge (if not above) its range of
applicability to lattice data where finite size effects start to play a role.

In fact, up to $m_\pi\lesssim(500-550)\,{\rm MeV}$ the $\chi$PT to
${\cal O}(p^4)$ \cite{AliKhan:2003cu,Procura:2003ig,Bernard:2005fy} and the 
$\chi$QSM describe $M_N(m_\pi)$ in good qualitative agreement, see Fig.~\ref{Fig-4}a.
Beyond this point, however, the nucleon mass as function of $m_\pi^2$ from
$\chi$PT in Refs.~\cite{AliKhan:2003cu,Procura:2003ig} changes the curvature,
indicating that the range of reliability of $\chi$PT to ${\cal O}(p^4)$ 
could be $m_\pi \lesssim 500\,{\rm MeV}$ (which, in fact,
is not far from the generally assumed bound $m_\pi < 600\,{\rm MeV}$).
This contrasts the $\chi$QSM-result exhibiting in agreement with lattice
data negative curvature up to the highest considered $m_\pi^2$.

With the above considerations in mind, we conclude that the $\chi$QSM describes
lattice data \cite{AliKhan:2001tx,Aoki:2002uc,Allton:2001sk,AliKhan:2003cu}
constrained by $a < 0.15\,{\rm fm}$, $m_\pi L > 5$ and $L \ge 2\,{\rm fm}$ very
well, see Fig.~\ref{Fig-4}a and c.f.\ Fig.~\ref{Fig-NEW}. 
In the range below $m_\pi\lesssim 500\,{\rm MeV}$
we observe a good qualitative and quantitative agreement of the $\chi$QSM
with $\chi$PT to ${\cal O}(p^4)$ \cite{AliKhan:2003cu,Procura:2003ig,Bernard:2005fy}.
In this range the curves for $M_N(m_\pi)$ from the two approaches agree
with each other to within an accuracy of $50\,{\rm MeV}$ and better, see 
Fig.~\ref{Fig-NEW} and \ref{Fig-4}a. This number may give us a flavour of the 
magnitude of the systematic error of the $\chi$QSM-extrapolation of lattice data,
though $\chi$PT to ${\cal O}(p^4)$ may also have such an intrinsic uncertainty,
as indicated in Table~\ref{Table:CQSM-vs-CPT}, due to unestimated
contributions from ${\cal O}(p^5)$.

%
\begingroup
\squeezetable
\begin{table}[b!]
    \caption{\footnotesize\sl
    \label{Table:CQSM-vs-CPT}
  The value of the physical nucleon mass $M_N$
  as extrapolated by means of $\chi$PT to ${\cal O}(p^4)$ lattice data
  \cite{AliKhan:2001tx,Aoki:2002uc,Allton:2001sk,AliKhan:2003cu} subject to
  the constraints $a < 0.15\,{\rm fm}$, $m_\pi L > 5$ and respectively
  $m_\pi<800\,{\rm MeV}$ in Ref.~\cite{AliKhan:2003cu}, and
  $m_\pi<600\,{\rm MeV}$ in Ref.~\cite{Procura:2003ig}.
  For comparison the $\chi$QSM-fits to the same data set, and to the same
  data set subject to the additional condition $L\ge 2\,{\rm fm}$ are shown.
  The error of $M_N$ due to the statistical uncertainty of the lattice data
  and the $\chi^2_{\rm dof}$ of the fit are shown.
  The systematic error of $M_N$ due to the extrapolation method, as well as
  lattice discretization or finite size effects, is merely indicated.
  (See the remark in footnote~\ref{footnote:accuracy} concerning the small
  error of the $\chi$QSM-fit due to the statistical uncertainty of lattice
  data.)}
\vspace{0.2cm}
    \begin{ruledtabular}
    \begin{tabular}{llr}
    \\
    Method & $M_N$ in MeV & $\chi^2_{\rm dof}$\\
    \\
    \hline
    \\
$\chi$PT to ${\cal O}(p^4)$, ``Fit I'' in Ref.~\cite{AliKhan:2003cu}  &
$948 \pm 60 \pm {\rm syst}$ &  1.7\\
$\chi$PT to ${\cal O}(p^4)$, ``Fit II'' in Ref.~\cite{Procura:2003ig} &
938 (fixed) & -- \\
$\chi$QSM (only $L\gtrsim 2\,{\rm fm}$)& $927 \pm 4 \pm {\rm syst}$ &  0.5   \\
$\chi$QSM (all data points)            & $931 \pm 4 \pm {\rm syst}$ &  2.3  \\
    \\
\end{tabular}
\end{ruledtabular}
\end{table}
\endgroup
%

Next we consider the 
effective approach based on the finite range regulator (FRR) method of 
Ref.~\cite{Leinweber:2003dg}. There 5 free parameters appear, which can well
be constrained by lattice data thanks to the larger range of applicability of
the approach, namely up to $m_\pi^2=1\,{\rm GeV}^2$.
In Ref.~\cite{Leinweber:2003dg} different shapes of regulators were exploited
and shown to yield practically the same results. Fig.~\ref{Fig-4}b shows
the 5-parameter fit (using the dipole-type regulator) to the CP-PACS lattice
data sets with $\beta=2.1$ and $2.2$ from Ref.~\cite{AliKhan:2001tx}.
In the $\chi$QSM-approach with one free parameter only, we were able to
fit both data sets separately, see Figs.~\ref{Fig-2}c and d. For sake of
comparison we include the combined $\chi$QSM-fit in Fig.~\ref{Fig-4}b
and summarize the results in Table~\ref{Table:CQSM-vs-FRR}.

As demonstrated in Fig.~\ref{Fig-4}b and Table~\ref{Table:CQSM-vs-FRR}
the results for $M_N(m_\pi)$ from the FRR approach and the $\chi$QSM
describe the CP-PACS data \cite{AliKhan:2001tx} equally well.
Having a closer look on the region $m_\pi^2<0.3\,{\rm GeV}^2$ we
see that $M_N(m_\pi)$ from the $\chi$QSM starts to deviate more and
more strongly from the fit of the FRR-approach with decreasing $m_\pi^2$,
though the curves remain very similar.
The reason for the discrepancy (which apparently was of no relevance for 
$m_\pi^2 > 0.3\,{\rm GeV}^2$, see Fig.~\ref{Fig-4}b) should be attributed
to the fact, that $M_\Delta-M_N$ is kept finite in the FRR approach, but
neglected in the $\chi$QSM.

%
\begingroup
\squeezetable
\begin{table}[t!]
    \caption{\footnotesize\sl
    \label{Table:CQSM-vs-FRR}
  The value of the physical nucleon mass $M_N$ as extrapolated from CP-PACS
  lattice data \cite{AliKhan:2001tx} obtained from lattices of the sizes
  $L=(2.2-3.1)\,{\rm fm}$ with lattice spacings $a=(0.09-0.13)\,{\rm fm}$
  covering the range $520\,{\rm Mev} \le m_\pi\le 970\,{\rm MeV}$.
  As guidelines for the extrapolation were used the finite range regulator
  approach \cite{Leinweber:2003dg} and the $\chi$QSM. See also the remarks
  in the caption to Table~\ref{Table:CQSM-vs-CPT}.}
\vspace{0.2cm}
    \begin{ruledtabular}
    \begin{tabular}{llr}
    \\
    Method & $M_N$ in MeV & $\chi^2_{\rm dof}$\\
    \\
    \hline
    \\
finite range (``dipole'') regulator \cite{Leinweber:2003dg}
       & $959 \pm 116 \pm {\rm syst}$ &  0.4  \\
$\chi$QSM  & $920 \pm   3 \pm {\rm syst}$ &  1.2  \\
    \\
\end{tabular}
\end{ruledtabular}
\end{table}
\endgroup
%

Thus one is lead to the conclusion, that the FRR method and the $\chi$QSM 
are completely consistent modulo $1/N_c$ corrections.

At the physical point the difference between the values of $M_N$ extrapolated
by means of the FRR method and the $\chi$QSM is about $40\,{\rm MeV}$, which
again may give us a rough idea on the theoretical uncertainty due to
neglecting the $\Delta$-nucleon mass difference.
Note that for this rough estimate we use the central value of the
extrapolated $M_N$ from the FRR approach, see Table~\ref{Table:CQSM-vs-FRR},
which has a substantially larger statistical uncertainty arising from
fitting 5 parameters to the lattice data. In this respect the 
$\chi$QSM-fit is far more precise.

To summarize, the comparison of the $\chi$QSM-based fits and those obtained 
using the first-principle approach in $\chi$PT 
\cite{AliKhan:2003cu,Procura:2003ig} and the FRR approach
\cite{Leinweber:2003dg} leads us to the following conclusion.
The systematic uncertainty of $M_N$ due to neglecting the finite
$\Delta$-nucleon mass-splitting in the $\chi$QSM is effectively of the order
of magnitude of $50\,{\rm MeV}$ with the tendency to underestimate
the nucleon mass. Noteworthy, a similar result -- concerning both sign and 
order of magnitude -- follows from a crude estimate within the model itself, 
see App.~\ref{App-C}.

Recall that we did not consider isospin breaking effects or
electromagnetic corrections $\sim {\cal O}(10\,{\rm MeV})$, 
see footnote~\ref{footnote:accuracy}.
Including this we have to assign a systematic error to the $\chi$QSM-fits 
of about $(\delta M_N)_{\rm syst} \approx 60\,{\rm MeV}$. 
Taking into account a systematic error of this magnitude we observe 
that the extrapolated values in Tables~\ref{Table:Mn-extrapolated}, 
\ref{Table:CQSM-vs-CPT} and \ref{Table:CQSM-vs-FRR} are all consistent 
with the physical mass of the nucleon.

We stress that 
we do not see any possibility to quantify the uncertainty of 
the $\chi$QSM-extrapolation of lattice data more quantitatively than that.

\section{Conclusions}
\label{Sec:5-conclusions}

The implicit dependence of the nucleon mass $M_N$ on the pion
mass $m_\pi$ was studied in the large-$N_c$ limit in the framework of the
chiral quark soliton model. 
The $M_N(m_\pi)$ in the model exhibits a chiral behaviour and includes
leading non-analytic terms which are consistent with the large-$N_c$
formulation of QCD \cite{Schweitzer:2003sb}.
As was shown here, the model describes correctly also the heavy quark limit.
The most remarkable observation we make here is that the model results 
for $M_N(m_\pi)$ well describe lattice data from full simulations
\cite{Aoki:1999ff,AliKhan:2001tx,Bernard:2001av,Aoki:2002uc,Allton:2001sk,Aoki:2004ht,AliKhan:2003cu}
over the wide range of pion masses
$0.1\,{\rm GeV}^2 < m_\pi^2 < 2.5\,{\rm GeV}^2$,
provided one takes into account the generic overestimate of the nucleon
mass in the soliton approach \cite{Pobylitsa:1992bk}. This is done by
introducing an $m_\pi$-independent subtraction constant, i.e.\
one single parameter to be fitted to the respective lattice data set.

The good description of the lattice data on $M_N(m_\pi)$ can 
{\sl partly} be understood as follows. In the $\chi$QSM, 
in the leading order of the large-$N_c$ limit, the
$\Delta$-nucleon mass-splitting $\Delta=M_\Delta-M_N \sim {\cal O}(N_c^{-1})$ 
is neglected. That is a reasonable approximation when comparing
to the present day lattice simulations where $\Delta^2 \ll m_\pi^2$ holds.
However, the remarkable precision, to which the model describes the lattice 
data, remains a puzzle --- to be clarified by further model studies.

We observe that the values for the nucleon mass ``extrapolated'' from the 
lattice data \cite{Aoki:1999ff,AliKhan:2001tx,Bernard:2001av,Aoki:2002uc,Allton:2001sk,Aoki:2004ht,AliKhan:2003cu}
on the basis of results from the $\chi$QSM 
are in good agreement with extrapolations based on the first principle approach 
in $\chi$PT \cite{AliKhan:2003cu,Procura:2003ig} or the effective FRR approach of 
Ref.~\cite{Leinweber:2003dg}, and agree with the physical nucleon mass to within $5\%$.
(But one has to keep in mind that at the same time --- for the adopted
handling of model parameters in the chiral limit --- the lattice values of the pion 
decay constant at large $m_\pi\sim1\,{\rm GeV}$ are underestimated by up to
$40\,\%$.)

It is difficult to exactly quantify the theoretical uncertainty of this
extrapolation due to the model-dependence. The main limitation for using 
the $\chi$QSM as a  guideline for the chiral extrapolation of lattice data 
is due to the large-$N_c$ limit. There is no strict control within the model 
of the theoretical uncertainty introduced by neglecting the finite 
$\Delta$-nucleon mass-splitting at the physical point, and we cannot 
quantify this and other uncertainties due to model-dependence quantitatively.
This limits the use of the model as a {\sl quantitative} effective tool
for the extrapolation of lattice data.

Still, the model may provide interesting {\sl qualitative} insights --- 
in particular in those cases when the matching between lattice results 
and $\chi$PT is difficult. A prominent example for that are (moments of) 
structure functions. 
In order to use the $\chi$QSM as a qualitative, but within its model
accuracy reliable device, which is helpful for a comparison of lattice 
results to experimental data, further model studies are 
necessary --- concerning the issue of handling model parameters in the 
chiral limit as well as addressing other observables \cite{work-in-progress}.

From the model point of view the observations made in this work also 
contribute to a better understanding of the physics which underlies the 
chiral quark-soliton model, and may --- that is our hope --- stimulate 
further studies in this direction in this, and perhaps also other models.

\acknowledgements
We thank Dmitri Diakonov, Thomas Hemmert, Victor Petrov, 
Pavel Pobylitsa, Maxim Polyakov, Gerrit Schierholz, Wolfram Schroers
and Tony Thomas for fruitful discussions and valuable comments.

This research is part of the EU integrated infrastructure initiative
hadron physics project under contract number RII3-CT-2004-506078,
and partially supported by the Graduierten-Kolleg Bochum-Dortmund
and Verbundforschung of BMBF.
A.~S.\  acknowledges support from GRICES and DAAD.

\appendix
\section{Proper-time regularization}
\label{App:A-proper-time}

The proper-time regularized versions of the integrals $I_1$ and $I_2$
appearing in Eqs.~(\ref{Eq:vacuum-condensate},~\ref{Eq:fpi}) are given by
\ba\label{App-01}
    I_1(m) &=& \!\!
    \int\limits_{\Lambda_{\rm cut}^{-2}}^\infty\!\frac{\di u}{u^2}\;
    \frac{\exp(-u {M^\prime}^2)}{(4\pi)^2}\,, \\
    I_2(m)&=& \!\!
    \int\limits_{\Lambda_{\rm cut}^{-2}}^\infty\!\!\frac{\di u}{2 u}\;
    \frac{\exp(-u {M^\prime}^2)}{(4\pi)^2}\,
    \int\limits_0^1\di\beta \exp( u \beta(1-\beta) m_\pi^2)\nonumber
\ea
where the $m$-dependence is hidden in $M^\prime\equiv M + m$.
For a given $m_\pi$ and $f_\pi=93\,{\rm MeV}$ fixed and due to
Eqs.~(\ref{Eq:fpi},~\ref{Eq:mpi}) both the current quark mass and
the cutoff are (implicit) functions of $m_\pi$, i.e. $m=m(m_\pi)$
and $\Lambda_{\rm cut}=\Lambda_{\rm cut}(m_\pi)$; some selected
values are shown in Table~\ref{Table:dependence-on-mpi}.

The expression for the soliton energy (\ref{Eq:soliton-energy})
in the proper-time regularization is given by
\ba\label{App-02}
&&      E_{\rm sol} = N_c \biggl[E_{\rm lev}+
        \sum\limits_n \bigl(R(E_n)-R(E_{n_0})\bigr)\biggr]\;;\nonumber\\
&&      R(\omega) = \frac{1}{4\sqrt{\pi}}
        \int\limits_{\Lambda^{-2}_{\rm cut}}^\infty
        \!\!\frac{\di \, u}{u^{3/2}} \,\exp(-u\omega^2) \;.
\ea

\section{\boldmath The pion-nucleon sigma-term $\sigma_{\pi N}$}
\label{App:B-sigma}

The pion-nucleon sigma-term is an important quantity to learn about chiral
symmetry breaking effects in the nucleon. The Feynman-Hellmann theorem
\cite{Feynman-Hellmann-theorem} relates $\sigma_{\pi N}$ to the slope
of $M_N(m)$ (with $m=m_q=m_u=m_d$ neglecting isospin breaking effects)
as follows
\be\label{Eq:sigma-term}
    \sigma_{\pi N} \equiv m\,\frac{\partial M_N(m)}{\partial m}
    = m_\pi^2\,\frac{\partial M_N(m_\pi)}{\partial m_\pi^2}
\ee
where the second equality holds, strictly speaking, only for small $m_\pi$. 
Eq.~(\ref{Eq:sigma-term}) offers a convenient way to learn 
about $\sigma_{\pi N}$ from lattice calculations of $M_N(m_\pi)$, 
see e.g.\  \cite{Leinweber:2000sa}. Direct lattice calculations are more
difficult but possible, see e.g.\  \cite{Dong:1995ec}.

Apart from (\ref{Eq:sigma-term}) one also can evaluate in the model directly
the actual definition of $\sigma_{\pi N}$ as double commutator of the strong
interaction Hamiltonian with two axial isovector charges
\cite{Wakamatsu:1992wr,Kim:1995hu}, or exploit a sum rule for the twist-3
distribution function $e(x)$ \cite{Schweitzer:2003uy}. All three methods
yield the same result in the $\chi$QSM \cite{Schweitzer:2003sb}.

By means of Eq.~(\ref{Eq:sigma-term}) we obtain in the present study 
in the proper-time regularization: $\sigma_{\pi N}=40\,{\rm MeV}$.
This agrees well with other $\chi$QSM calculations 
\cite{Diakonov:1988mg,Wakamatsu:1992wr,Kim:1995hu}
performed in this regularization scheme.

In the $\chi$QSM the pion-nucleon sigma-term is quadratically
UV-divergent, and as such particularly sensitive to details of
regularization. In Ref.~\cite{Schweitzer:2003sb} a way was found
to compute $\sigma_{\pi N}$ in a regularization scheme independent
way with the result $\sigma_{\pi N}\approx 68\,{\rm MeV}$.
The prize to pay for the regularization scheme independence was the
use of an {\sl approximation} expected to work within an accuracy of
${\cal O}(30\%)$ and well-justified by notions from the instanton vacuum model.

On the basis of our results for $M_N(m_\pi)$ we are now
in a position to check the accuracy of this approximation in practice.
For that we note that the Feynman-Hellmann relation (\ref{Eq:sigma-term})
allows to determine $M_N(m_\pi)$ from $\sigma_{\pi N}(m_\pi)$ up to an
integration constant.

From the approximate (but regularization scheme independent) result for
$\sigma_{\pi N}(m_\pi)$ from Ref.~\cite{Schweitzer:2003sb} one obtains
for $M_N(m_\pi)$ the result shown as dashed line in Fig.~\ref{Fig-1}, where
for convenience the integration constant is chosen such that both curves
coincide at a central value of $m_\pi=1\,{\rm GeV}$ in the plot.
We observe a good agreement of our exact result and the approximation of
Ref.~\cite{Schweitzer:2003sb}, see Fig.~\ref{Fig-1}. 

Also our result for $\sigma_{\pi N}$ agrees with the 
regularization scheme-independent result for $\sigma_{\pi N}$ from 
Ref.~\cite{Schweitzer:2003sb} to within the expected accuracy of 
${\cal O}(30\%)$.

Thus, our results confirm that the instanton model motivated approximation
advocated in \cite{Schweitzer:2003sb} works well. Note, however, that
here we went far beyond the study of Ref.~\cite{Schweitzer:2003sb}.
In this work we practically demonstrated that stable soliton solutions
{\sl do exist} also for large values of the pion mass, while in
Ref.~\cite{Schweitzer:2003sb} this was {\sl presumed}.

The regularization scheme independent result of Ref.~\cite{Schweitzer:2003sb}
is in good agreement with recent extractions indicating
$\sigma_{\pi N}=(60-70)\,{\rm MeV}$ \cite{recent-sigma-extractions},
which is substantially more sizeable than the value obtained from earlier
analyses \cite{earlier-sigma-extractions}.

In this context it is worthwhile mentioning that the spectrum of exotic
("pentaquark") baryons which were predicted in the framework of the
$\chi$QSM \cite{Diakonov:1997mm} and for which recently possible
observations were reported -- see \cite{Hicks:2004ge} for recent overviews --
could provide an independent mean to access information on the pion-nucleon
sigma-term \cite{Schweitzer:2003fg}. The presently available data on the
exotic baryons favour a large value for $\sigma_{\pi N}$ as found in
\cite{recent-sigma-extractions}.

\section{Systematic uncertainty of the rotating soliton approach}
\label{App-C}

The neglect of $\Delta=M_\Delta-M_N$ in the leading order of the large-$N_c$
limit introduces a systematic uncertainty which is difficult to quantify.
Here we roughly estimate this uncertainty on the basis of the soliton approach.

In this approach the nucleon and the $\Delta$-resonance are just different
rotational excitations of the same classic object, the soliton. A non-zero
mass difference between the nucleon and the $\Delta$-resonance arises due
to considering a particular class of (``rotational'') $1/N_c$-corrections
\cite{Diakonov:yh,Diakonov:1987ty}. The mass of an SU(2)-baryon $M_B$
($B=N$ or $\Delta$ in the real world with $N_c= 3$ colours)
with spin $S_B$ is given by
\be
    M_B = E_{\rm sol}+M_2+S_B(S_B+1)\,\frac{\Delta}{3} + \dots
\ee
with $E_{\rm sol}$ as defined in Eq.~(\ref{Eq:soliton-energy}), $M_2$ denoting
as in Eq.~(\ref{Eq:Mn-in-large-Nc}) the ${\cal O}(N_c^0)$ correction to the
baryon mass which is the same for the nucleon and the $\Delta$-resonance,
and the dots representing higher order $1/N_c$-corrections.

Then we obtain as a first correction to $M_N$ in
Eq.~(\ref{Eq:Mn-as-funct-of-mpi-Delta-2})
\be
    M_N(m_\pi,\Delta) = M_N(m_\pi,0) + \frac{\Delta(m_\pi)}{4} + \dots
\ee
Focusing on the linear order correction in $\Delta$ and neglecting higher
orders, we see that by neglecting $\Delta$ one underestimates the nucleon mass
by about $70\,{\rm MeV}$ at the physical value of the pion mass. This is
in good agreement with the systematic uncertainty roughly estimated in
Sec.~\ref{Sec:4-using-for-extrapolation?}.


\end{document}